\documentclass[%
 reprint,
 amsmath,amssymb,
 aps,pra,superscriptaddress
]{revtex4-2}
\usepackage{graphicx}% Include figure files
\usepackage{dcolumn}% Align table columns on decimal point
\usepackage{bm}% bold math
\usepackage{multirow}
\usepackage{makecell}
\usepackage{array}
\usepackage{hyperref}% add hypertext 
\hypersetup{colorlinks=true,linkcolor=blue,filecolor=blue,urlcolor=blue,citecolor=blue,}
\usepackage[mathlines]{lineno}% Enable 

\begin{document}
\preprint{APS/123-QED}

\title{General Criteria for Certifying Genuine High-Dimensional Quantum Teleportation}

\author{Neng-Fei Gong}
\affiliation{State Key Laboratory of Information Photonics and Optical Communications and School of Physical Science and Technology, Beijing University of Posts and Telecommunications, Beijing 100876, China}

\author{Tie-Jun Wang}
\email{\textcolor{black}{wangtiejun@bupt.edu.cn}}
\affiliation{State Key Laboratory of Information Photonics and Optical Communications and School of Physical Science and Technology, Beijing University of Posts and Telecommunications, Beijing 100876, China}

\begin{abstract}

Developing reliable methods for certifying the dimension of a given quantum system or process is essential to ensure the validity of claimed realization of high-dimensional (HD) quantum advantages. The existing criteria for certifying genuine HD quantum teleportation (HDQT) mainly focus on demonstrating the successful transmission of genuine HD quantum states. However, a complete certification of HDQT must also identify the entanglement dimension of resource, which is critical for verifying whether the transmission capacity and noise resilience meet the necessary thresholds. Here we propose two universal criteria (based on fidelity and robustness, respectively) for certifying genuine HDQT behaviors that can close this gap by fully identifying the dimension of the entanglement. Both criteria require only the input and output teleportation data and remain feasible under partial Bell-state measurements. Furthermore, the robustness-based criterion has stronger noise resistance and it requires no prior assumptions about local operations, making it robust even in black-box scenario. Our results establish a universal and reliable theoretical framework for validating the core quantum advantage in HDQT, pivotal for ensuring the reliable links in HD quantum networks.

\end{abstract}
  
\maketitle

\section{\label{sec:1}Introduction}
Quantum teleportation (QT) \cite{bennett1993teleporting,pirandola2015advances,hu2023progress}, a technique that enables the transmission of an unknown quantum state from one location to another without the physical movement of quantum information carriers, serves as a cornerstone of distributed quantum computing and communicating networks \cite{dias2017quantum,pompili2021realization,hermans2022qubit,liu2024creation}. While the QT of a qubit has been extensively demonstrated \cite{bouwmeester1997experimental,pan2001experimental,li2022quantum,yin2012quantum,ma2012quantum,ren2017ground,gonzalez2024satellite,lago2023long}, high-dimensional QT (HDQT) has recently gained attention \cite{luo2019quantum,hu2020experimental,liu2024deterministic} because of its superior information capacity and enhanced resilience against noise in physical carriers such as optical systems \cite{cozzolino2019high,liu2020orbital,erhard2020advances,bacco2021proposal}. To ensure the reliability of the claimed realization of HD quantum advantages, it is essential to develop reliable methods for certifying whether a HDQT process has at least a certain dimension i.e. certifying genuine HDQT \cite{luo2019quantum,hu2020experimental,liu2024deterministic} .

The existing methods for certifying genuine HDQT proposed in Ref. \cite{luo2019quantum,hu2020experimental} focus on demonstrating the successful teleportation of a genuine qutrit. Based on which the authors proposed a fidelity criterion $F^{\psi_0}_{\text {tel }}>2/3$ where $F^{\psi_0}_{\text {tel }}$ is the measured teleportation fidelity when the input state is $|\psi_0\rangle=(|0\rangle+|1\rangle+|2\rangle)/\sqrt{3}$ \cite{luo2019quantum}, and a robustness criterion $\mu>0$, where $\mu$ is the minimum amount of “white noise” that must be added to the qutrit state such that the mixture can be simulated by qubit states \cite{hu2020experimental}. Such criteria are useful for experimenters to verify that the setup is capable of teleporting genuine HD quantum states.

However, the certification of genuine HDQT has a twofold objective: it must assess not only the ability to teleport genuine HD quantum states, but also whether the process employs HD entanglement \cite{horodecki2009quantum, brunner2014bell,uola2020quantum}. The presence of $d$-dimensional entanglement provides the necessary assurance that the devices are capable of perfectly teleporting arbitrary states from $d$-dimensional Hilbert space (in ideal case), which is essential in broader scenarios where the state to teleport is assumed as unknown. Despite transmission capacity, employing HD entanglement also ensures that other HD quantum advantages meet the necessary thresholds, such as the robustness against noise and attack  \cite{erhard2020advances}. 

Previous criteria \cite{luo2019quantum,hu2020experimental}, as we show in Appendix ~\ref{Ap:A}, cannot be employed to identify the entanglement dimension used in HDQT process, since they can be satisfied by merely lower-dimensional entanglement with local operations. Certifying the entanglement dimension of the channel state  \cite{martin2017quantifying, bavaresco2018measurements,wyderka2023construction,lib2025experimental} before teleportation process is a potential solution, but it is unfeasible in practical scenarios where the verifier cannot preemptively access the channel state or the teleportation devices cannot be assumed as trusted. A universal certification should be executed solely on analyzing input-output state pairs that constitute the teleportation data \cite{cavalcanti2017all}. Hence, it is necessary to establish a criterion to test whether a set of HD teleportation data can be simulated using a channel state of lower-dimensional entanglement. We refer to such teleportation data that cannot be simulated by lower-dimensional entanglement as exhibiting a genuine HDQT behavior. Furthermore, to ensure broad applicability, one should go a step further and precisely identify the actual dimension of the HDQT behavior \cite{key}.

In this paper, we address this gap by proposing two universal criteria using teleportation data for certifying genuine $d^\prime$-dimensional QT behaviors when teleporting $d$-dimensional states ($d^\prime \leq d$). The first one is based on average teleportation fidelity,  $\bar{F}_{\mathrm{tel}}>d^\prime/(d+1)$. Satisfying this criterion indicates that the process exhibits at least genuine $d'$-dimensional QT behavior. Although fidelity is the most common metric in quantum information processing, it still has shortcomings in terms of verification, such as limited noise resistance \cite{cobucci2024detecting} and dependence on optimal local operations for achieving optimal average fidelity \cite{gong2024optimal}. Hence, we go beyond fidelity and propose a robustness-based criterion which has stronger noise resistance and is feasible without any assumption of local operations. We introduce the generalized genuine $d'$-dimensional QT robustness ($\text{GHTR}_{d^\prime}$), which is defined by the minimum amount of the generalized noise that has to be added to the teleportation data such that the teleportation data can be simulated with channel state of  only $(d^\prime-1)$-dimensional entanglement. Any positive value of $\text{GHTR}_{d^\prime}$ certifies at least genuine $d^\prime$-dimensional QT behavior. Both the proposed criteria are feasible with partial Bell-state measurements (BSM), and the robustness-based criterion exhibits stronger certification capability as it can identify those ‘weak’ HDQT while the fidelity-based criterion fails. We go further and prove that, with the robustness-based criterion, all channel states with $d'$-dimensional entanglement enable genuine $d'$-dimensional QT behaviors.

The paper is organized as follows. In Sec.~\ref{sec:2}, the certification protocol of HDQT is briefly given. In Sec.~\ref{sec:3}, the fidelity-based criterion and its corresponding application in characterizing the hierarchy of HDQT performance are presented in Sec.~\ref{sec:3:A} and Sec.~\ref{sec:3:B}, respectively. In Sec. ~\ref{sec:4}, the details about the robustness-based criterion is given in Sec. ~\ref{sec:4:A}, its application in demonstrating that all channel states with $d'$-dimensional entanglement enable genuine $d'$-dimensional QT behavior is given in Sec. ~\ref{4:B}, and its advantage in noise resilience is discussed in Sec. ~\ref{4:C}. Finally, a discussion and summary are given in Sec.~\ref{5}. Detailed derivations and supplementary examples are provided in the Appendices.”

\section{\label{sec:2}Certification protocol of quantum teleportation}

\begin{figure}[h!]
\centering
\includegraphics[width=8cm]{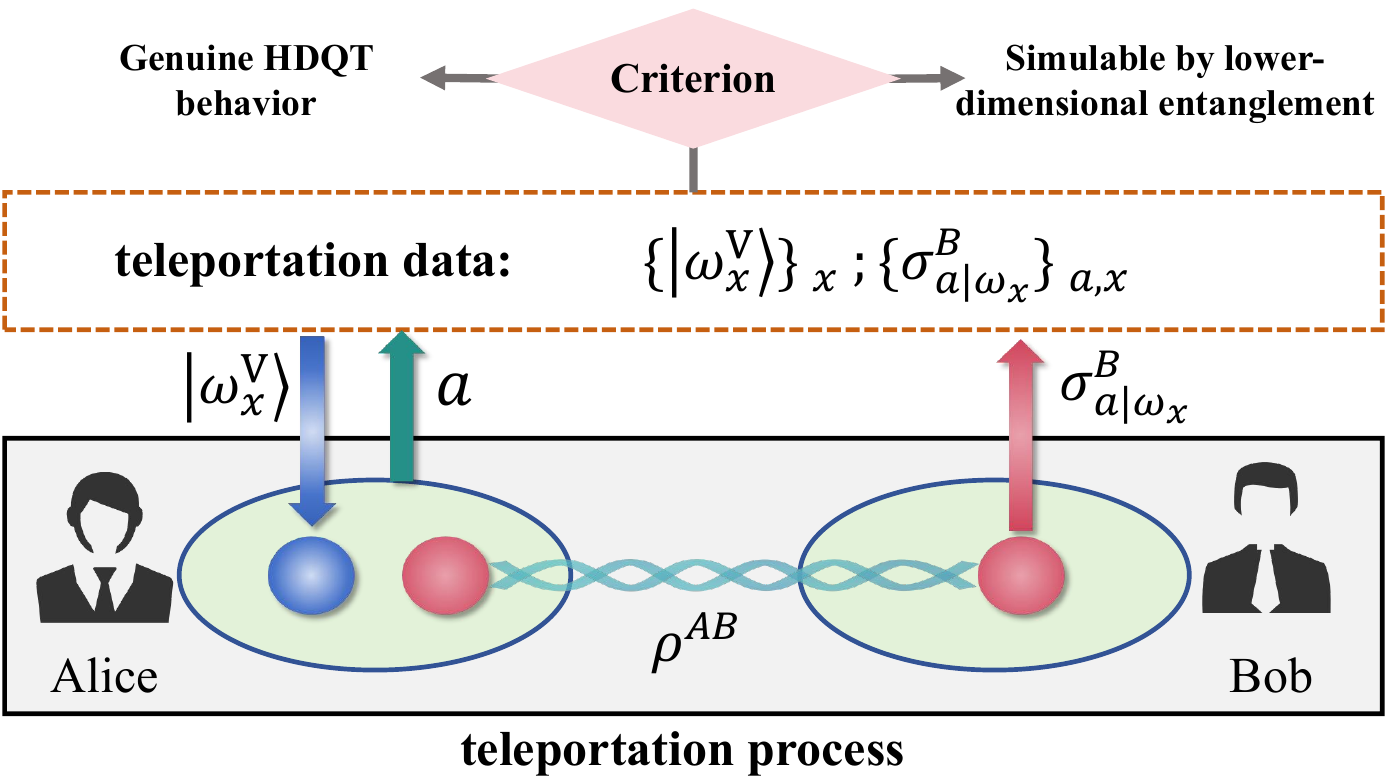}
\caption{Certification protocol of quantum teleportation. Alice and Bob share a bipartite quantum state and they claim that they can supply a HDQT service between themselves. Charlie needs the service and to certify the quality of the teleportation. He sends a set of states $\{\left|\omega^{\mathrm{V}}_x\right\rangle\}_x$ (unknown to Alice and Bob) to Alice, and receives output states $\{\sigma^{\mathrm{B}}_{a|\omega_x}\}_{a,x}$ corresponding to Alice's readout $a$. Using these input and output data, Charlie can execute some criterion to provide whether to accept or reject.}
\label{fig: QT scheme}
\end{figure}

Suppose two spatially separated parties, Alice and Bob, as depicted in Fig. \ref{fig: QT scheme}, share a bipartite quantum state and they claim that they are capable of performing $d$-dimensional QT. To verify that the claim is true, a third party Charlie (the verifier) provides quantum systems V to Alice in pure states $\left|\omega^{\mathrm{V}}_x\right\rangle$, $x=1, \ldots,|x|$  (the state set $\{|\omega^{\mathrm{V}}_x\rangle\}_x$ is  distributed uniformly or tomographically complete in $d$-dimensional pure-state region), which are unknown to Alice, and asks her to teleport these states to Bob. After the teleportation completed, Bob sends his system to the verifier, who is able to check whether it is the same as the one provided to Alice. In ideal $d$-dimensional QT protocol, the channel state shared is one of the $d$-dimensional Bell states, $e.g. \left|\phi^{\mathrm{AB}}\right\rangle=(1/\sqrt{d})\sum_{i=0}^{d-1}|i^\mathrm{A}\rangle|i^\mathrm{B}\rangle$. Then Alice perform a joint $d\times d$-dimensional Bell-state measurement on systems V and A. This measurement is carried out in the generalized Bell basis, which for two qudits comprises $d^2$ maximally entangled states:
\begin{equation}
\label{eq:bell_basis}
    |\Phi_{mn,d}^{\text{V} \text{A}}\rangle = \frac{1}{\sqrt{d}} \sum_{l=0}^{d-1} e^{2\pi \text{i} l m / d} |l^{\text{V}}\rangle |(l+n)^{\text{A}}\rangle,
\end{equation}
where $m\in \{0, 1, \dots, d-1\}$ represents the relative phase information and $n\in \{0, 1, \dots, d-1\}$ ($l+n$ must be taken from modulo $d$). $\text{i}$ is the imaginary unit. Through the measurement, Alice projects Bob's subsystem B into the state $\rho_{mn \mid\omega_x}^{\mathrm{B}}=U^{\dagger}_{mn}\left|\omega_x\right\rangle\left\langle\omega_x\right| U_{mn}$, where $U^{\dagger}_{mn}$ is a known unitary transformation that corresponds to Alice's measurement outcome $mn$. Subsequent classical transmission of this outcome to Bob enables him to reverse the induced transformation by applying $U_{mn}$. It is completely equivalent if Alice previously shares the outcome $mn$ to the verifier instead of Bob, who can then check whether the state sent by Bob – which now remains uncorrected – matches the original state given to Alice, up to the known correction $U_{mn}$. 

Note that in the case of loophole-free verification, Charlie should ask Bob to transmit subsystem B to him before he sends state $\left|\omega^{\mathrm{V}}_x\right\rangle$ to Alice, hence to prevent Alice directly sending state  $U^{\dagger}_{mn}\left|\omega^{\mathrm{V}}_x\right\rangle\left\langle\omega^{\mathrm{V}}_x\right| U_{mn}$ to Bob, which is a transformation of input system V that corresponds to the broadcast information $mn$. 

In practical implementation of QT, the measurements are inevitably affected by noise and device imperfections, reducing Alice's projection into a set of POVM elements $\{M^{\mathrm{VA}}_a\}_a$ \cite{cavalcanti2017all}. On the other hand, Alice and Bob may try to simulate ideal $d$-dimensional QT by using other resources, such as channel state with lower-dimensional entanglement, denoted as  $\rho^{\mathrm{AB}}$ (Without loss of generality, $\rho^{\mathrm{AB}}$ is arbitrary bipartite state encoded in Hilbert space $\mathbb{H}_d^{\text{A}}\otimes \mathbb{H}_d^{\text{B}}$). In this case, the unnormalized output states that Bob receives when input state is $\left|\omega^{\mathrm{V}}_x\right\rangle$ and Alice's outcome $a$ can be written as
\begin{equation}
\sigma_{a \mid \omega_x}^{\mathrm{B}}=\operatorname{Tr}_{\mathrm{VA}}\left[(M_a^{\mathrm{VA}} \otimes \mathbb{I}^{\mathrm{B}}) \cdot(|\omega_x^{\mathrm{V}}\rangle\langle\omega_x^{\mathrm{V}}| \otimes \rho^{\mathrm{AB}})\right].
\label{rhoB}
\end{equation}
The input set $\{\left|\omega^{\mathrm{V}}_x\right\rangle\}_x$ and output set $\{\sigma_{a \mid \omega_x}^{\mathrm{B}}\}_{a, x}$ constitute the teleportation data, which the verifier has access to. These data are the only information Charlie needs to certify the teleportation process; he does not require direct access to the channel state or the internal workings of Alice’s and Bob’s devices. This “black‑box” approach is essential for practical certification of high‑dimensional quantum teleportation.

The goal of the certification protocol is to certify whether a given set of teleportation data, collected from a process of teleporting $d$-dimensional states, demonstrates a capability that fundamentally requires HD entanglement. We formalize this as follows:

\emph{Definition}-A set of teleportation data is said to exhibit genuine $d'$-dimensional QT behavior ($d'\leq d$) if and only if the data cannot be simulated by any process that uses a channel state whose entanglement dimension is at most $d'-1$.

\section{\label{sec:3}Certifying genuine HDQT behaviors via teleportation fidelity}

Having established the general certification protocol, we now develop our first concrete criterion. We begin with the fidelity-based approach because it is the most intuitive and widely used figure of merit in quantum information. In this section, we will: (i) derive the fidelity threshold that signals genuine $d$-dimensional teleportation, and (ii) show how this threshold naturally leads to a hierarchical classification of teleportation performance.

\subsection{\label{sec:3:A}Fidelity-based criterion}

The standard merit to evaluate the QT performance is the optimal average teleportation fidelity between the input and output states
\begin{equation}
\bar{F}_{\text{tel}}=\max _{\{U_a\}_a}\frac{1}{|x|} \sum_{a, x} \left\langle\omega_x\right| U_a \sigma_{a \mid \omega_x}^B U_a^{\dagger}\left|\omega_x\right\rangle .
\label{def.F}
\end{equation}
The maximum is obtained by choosing the optimal unitary operations $\{U_a\}_a$ which is determined by the knowledge of channel state and Alice's outcome $a$ \cite{gong2024optimal}. 

To certify if an experimentally-measured $\bar{F}_{\text{tel}}$ exhibits at least genuine $d^\prime$-dimensional QT behavior of teleporting $d$-dimensional state ($d^\prime\leq d$), one has to determine whether $\bar{F}_{\text{tel}}$ exceeds $\bar{F}^{d^\prime-1}_{q,d}$, which refers to the maximal average teleportation fidelity obtained by using channel state whose entanglement dimension is $d^\prime-1$ \cite{key}. Without loss of generality, we assume the simulation is performed via a single copy of $(d^\prime-1)$-dimensional entanglement. The entanglement dimensionality can be characterized by the concept of Schmidt number \cite{terhal2000schmidt}. The Schmidt number of state $\rho^{\mathrm{AB}}$ is the minimum $n$ such that there exists a decomposition $\rho^{\mathrm{AB}}=\sum_l p_l|\psi^\mathrm{AB}_l\rangle\langle\psi^\mathrm{AB}_l|$ where all $|\psi^\mathrm{AB}_l\rangle$ are pure entangled states of Schmidt rank at most $n$. Under this decomposition, Eq. \eqref{rhoB} can be written as $\sigma_{a \mid \omega_x}^{\mathrm{B}}=\sum_l p_l \sigma_{a \mid \omega_x}^{\mathrm{B},l}$, 
where $\sigma_{a \mid \omega_x}^{\mathrm{B},l}=\operatorname{Tr}_{\mathrm{VA}}[(M_a^{\mathrm{VA}} \otimes \mathbb{I}^{\mathrm{B}}) \cdot(|\omega^{\mathrm{V}}_x\rangle\langle\omega^{\mathrm{V}}_x| \otimes |\psi^\mathrm{AB}_l\rangle\langle\psi^\mathrm{AB}_l|)]$ is the output state corresponding to channel $|\psi^\mathrm{AB}_l\rangle\langle\psi^\mathrm{AB}_l|$. Using the concavity of fidelity function \cite{liang2019quantum}, we have
\begin{equation}
\begin{aligned}
\bar{F}_{q, d}^{d^{\prime}-1}\left(\rho^{\mathrm{AB}}\right)=&\sum_l p_l \bar{F}_{q, d}^{d^{\prime}-1}\left(\left|\psi_l^{\mathrm{AB}}\right\rangle\left\langle\psi_l^{\mathrm{AB}}\right|\right)\\ &\leq \bar{F}_{q, d}^{d^{\prime}-1}\left(\left|\psi_{l^{\prime}}^{\mathrm{AB}}\right\rangle\left\langle\psi_{l^{\prime}}^{\mathrm{AB}}\right|\right),
\label{decomp of fidelity}
\end{aligned}
\end{equation}
where $\left|\psi_{l^{\prime}}^{\mathrm{AB}}\right\rangle$ represents the pure channel state whose Schmidt rank is $d^\prime-1$. Eq. \eqref{decomp of fidelity} reflects a fundamental physical insight: if the channel state is a probabilistic mixture of several entangled pure states, the overall teleportation performance is at most as good as that of the component with the highest fidelity. This allows us to focus, without loss of generality, on pure channel states when deriving the upper bound for fidelity. According to Ref. \cite{gong2024optimal}, for a given channel state, the optimal average fidelity of teleporting $d$-dimensional states has a form of 
\begin{equation}
\bar{F}_{q, d}^{d^{\prime}-1}\left(\left|\psi_{l^{\prime}}^{\mathrm{AB}}\right\rangle\left\langle\psi_{l^{\prime}}^{\mathrm{AB}}\right|\right)=\max _O\left\{\frac{1}{d}+\frac{d}{4(d+1)} \operatorname{Tr}\left(T^d_{\phi} T_{\rho} O\right)\right\},
\label{fidelity and correlation matrix}
\end{equation}
where $T_{\rho}$, which refers to the correlation matrix of $\left|\psi_{l^{\prime}}^{\mathrm{AB}}\right\rangle$, is a $(d^2-1)$-dimensional square matrix whose elements are defined by $t^{\rho}_{ij}=\operatorname{Tr}(\left|\psi_{l^{\prime}}^{\mathrm{AB}}\right\rangle\left\langle\psi_{l^{\prime}}^{\mathrm{AB}}\right|\lambda_i \otimes \lambda_j)$, $\{\lambda_i\}^{d^2-1}_{i=1}$ is the set of the $d$-dimensional generalized Gell-Mann matrices \cite{bertlmann2008bloch}. $T^d_{\phi}$ refers to the correlation matrix of Bell state $\left|\phi_d\right\rangle=\sum_{i=0}^{d-1}(1/\sqrt{d})|i\rangle|i\rangle$. Eq. \eqref{fidelity and correlation matrix} is maximized over all possible rotation $O$, which refers to the homomorphic matrix of a global local unitary operation $U$ performed by Bob (by global we mean that $U$ is fixed under all possible outcome $a$), For a given $U$, the elements of $O$ can be determined via $o_{ij}= \operatorname{Tr}\left(U \lambda_i U^{\dagger} \lambda_j\right)/2, \quad  \forall i,j=1,2,\ldots, d^2-1$ \cite{li2008upper}. By derivation (see Appendix ~\ref{AP:B} for details), Eq. \eqref{fidelity and correlation matrix} can be simplified as 
\begin{equation}
    \bar{F}_{q, d}^{d^{\prime}-1}=\frac{d^\prime}{d+1}.
    \label{fmax of d'-1}
\end{equation}
Hence, we conclude that, when teleporting $d$-dimensional states, at least genuine $d^\prime$-dimensional QT behavior can be certified when the following criterion is satisfied,
\begin{equation}
\bar{F}_{\operatorname{tel}}>\frac{d^\prime}{d+1}.
\label{result 1}
\end{equation}
For instance, the fidelity-based criterion for genuine $3$-dimensional QT of teleporting a qutrit is $\bar{F}_{\operatorname{tel}}>3/4$. To illustrate that our result does not conflict with the basic principle of QT, we also present the corresponding three-dimensional QT scheme of obtaining average fidelity $3/4$ with two-dimensional Bell state (see Appendix \ref{Ap:D} for details). Furthermore, if $d^\prime=2$, Eq. \eqref{result 1} becomes $\bar{F}_{\operatorname{tel}}>2/(d+1)$, exactly the criterion for nonclassical teleportation \cite{pirandola2015advances}. 

As shown in Appendix ~\ref{Ap:E}, Eq. \eqref{result 1} is also feasible when Alice performs partial BSM, which could greatly reduce the cost of experimentally demonstrating genuine HDQT.

\subsection{\label{sec:3:B}Application: Characterizing the hierarchy of HDQT performance}

It is widely known that the range of achievable average fidelity of teleporting $d$-dimensional quantum states can be partitioned into two distinct regions: 
\begin{equation}
\begin{array}{ll}
\frac{2}{d+1}<\bar{F}_{\text {tel }}\leq 1 & \text{nonclassical region,} \\ \bar{F}_{\text {tel }}\leq \frac{2}{d+1} & \text{classical region.}
\end{array}
\end{equation}

Our proposed fidelity-based criterion for genuine HDQT behaviors establish a refined tiered classification for HDQT performance. According to Eq. \eqref{result 1}, when the average teleportation fidelity of the teleportation data satisfies $\bar{F}_{\text {tel }}\in (\frac{d'}{d+1}, \frac{d'+1}{d+1}]$, the teleportation data exhibits genuine $d'$-dimensional QT behavior of teleporting $d$-dimensional quantum states.

\begin{figure}[h!]
\centering
\includegraphics[width=8cm]{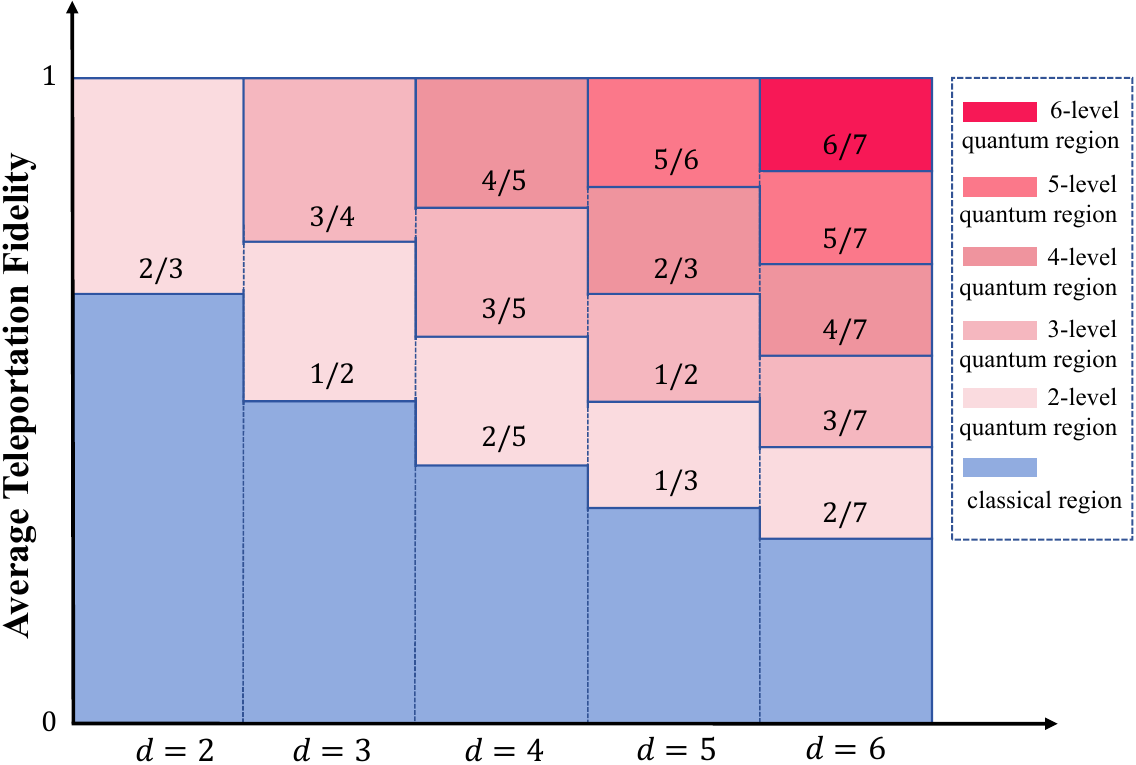}
\caption{The hierarchy of HDQT performance. For the case of teleporting $d$-dimensional quantum states, the region of achievable average teleportation fidelity can be divided into a classical region and $d-1$ quantum regions, including 2-level quantum region $(\frac{2}{d+1}, \frac{3}{d+1}]$,...,$d'$-level quantum region $(\frac{d'}{d+1}, \frac{d'+1}{d+1}]$.}
\label{Fig:hierarchy of HDQT performance}
\end{figure}

We refer $\Gamma^{d}_{d'}=(\frac{d'}{d+1}, \frac{d'+1}{d+1}]$ as $d'$-level quantum region in the case of teleporting $d$-dimensional quantum states. Following this idea, as shown in Fig. \ref{Fig:hierarchy of HDQT performance}, for given $d$, the nonclassical region can be divided into $d-1$ sub-regions, enabling a characterization of the hierarchy of HDQT performance. Driven by $d'$-dimensional entanglement, the HDQT behaviors in $d'$-level quantum region exhibit a step-like enhancement in both channel capacity and noise resistance compared to those in $(d'-1)$-level quantum region. This classification significantly enhances the granularity of performance evaluation for HDQT.

\section{\label{sec:4}Certifying genuine HDQT behaviors via semi-definite program}

To use the fidelity-based criterion, one has to experimentally measure the optimal teleportation fidelity with choosing the optimal set of local unitary operation $\{U_a\}_a$, which is hard to determine when accurate characterization of the noisy channel state is infeasible. Here we introduce a robustness-based criterion where no assumption about the local operations is required. And we show that this criterion outperforms the fidelity-based criterion in terms of detection ability.

\subsection{\label{sec:4:A}Robustness-based criterion}

The idea is to establish a convex program that can find the minimum amount of noises that has to be mixed with the collected teleportation data such that the mixture can be simulated by  any process that uses a channel state $\tilde{\rho}^{\text{A}\text{B}}$ whose entanglement dimension (i.e., the Schmidt number) is at most $d'-1$. 

To start, we denote the set of such states as $S_{d'-1}$. Usually, accurate characterization of $S_{d'-1}$ is challenging, and in many cases no analytical solution can be found. This can be remedied by relaxing the constraints slightly \cite{tavakoli2024semidefinite}. To this end, note that there exists a characterization of the set $S_{d'-1}$ in terms of positive maps \cite{terhal2000schmidt}: It holds that $\tilde{\rho}^{\mathrm{AB}}\in S_{d'-1}$ if and only if $(\mathbb{I}^{\mathrm{A}}\otimes \Lambda_{d'-1}^{\mathrm{B}})\tilde{\rho}^{\mathrm{AB}}\geq0$ for all ($d'-1$)-positive maps $\Lambda_{d'-1}^{\mathrm{B}}$. Such maps are positive when applied to one share of every bipartite state with a Schmidt number at most $(d^\prime-1)$ but non-positive for some states with a larger Schmidt number \cite{terhal2000schmidt}. This characterization is useful as it allows to define a slightly larger set than $S_{d'-1}$, which can be characterized with less effort and easy to be realized with linear programming. Hence, for all $\rho^{\mathrm{AB}}$ with Schmidt number at most $d^\prime-1$, the following constraint is satisfied,
\begin{equation}
\left(\mathbb{I}^\mathrm{A} \otimes \Lambda^\mathrm{B}_{d^\prime-1}\right)\tilde{\rho}^{\mathrm{AB}} \geq 0.
\label{Reduction map}
\end{equation}
Here, we use the generalized reduction map, $\Lambda_{d'-1}(\text{X})=\operatorname{Tr} (\text{X}) \mathbb{I}-\frac{1}{d'-1} \text{X}$, which is known to be a $d'-1$-positive map. This map is a commonly used positive map when dealing with Schmidt number of mixed states \cite{terhal2000schmidt,wyderka2023construction,cobucci2024detecting} and it was shown that the constraint is tight \cite{terhal2000schmidt}. Using this map, Eq. \eqref{Reduction map} is equal to
\begin{equation}
\text{Tr}_{\text{B}}[\tilde{\rho}^{\mathrm{AB}}]\otimes \mathbb{I}^{\text{B}}-\frac{1}{d'-1} \tilde{\rho}^{\mathrm{AB}} \geq 0
\label{Reduction map1}
\end{equation}
Next, we characterize the teleportation process based on Eq. \eqref{Reduction map1}. Suppose a set of collected teleportation data is $\{\omega_x\}_x$ (inputs, $\omega_x$ represents the density matrix of $|\omega_x\rangle$) and $\{\sigma^{\text{B}}_{{a|\omega_x}}\}_{a,x}$ (outputs). $\sigma^{\text{B}}_{{a|\omega_x}}$ has the structure as Eq. \eqref{rhoB}. To test how much amount of noise that has to be added onto $\sigma^{\text{B}}_{{a|\omega_x}}$ such that the mixture $\tilde{\sigma}_{a \mid \omega_x}^{\mathrm{B}}$ can be simulated with $\tilde{\rho}^{\mathrm{AB}}$, we write the overall output state as
\begin{equation}
\begin{aligned}
\tilde{\sigma}_{a \mid \omega_x}^{\mathrm{B}}=&\sigma_{a \mid \omega_x}^{\mathrm{B}}+\operatorname{Tr}_{\mathrm{V}}\left[N_a^{\mathrm{VB}}(|\omega_x^{\mathrm{V}}\rangle\langle\omega_x^{\mathrm{V}}| \otimes \mathbb{I}^{\mathrm{B}})\right]\\
=&\operatorname{Tr}_{\mathrm{V}}\left[M_a^{\mathrm{VB}}(|\omega_x^{\mathrm{V}}\rangle\langle\omega_x^{\mathrm{V}}| \otimes \mathbb{I}^{\mathrm{B}})\right],
\label{rhoBMvb}
\end{aligned}
\end{equation}
Compared with Eq. \eqref{rhoB}, operator  $M_a^{\mathrm{VB}}$ has the structure as $M_a^{\mathrm{VB}}=\operatorname{Tr}_{\mathrm{A}}\left[\left(M_a^{\mathrm{VA}} \otimes \mathbb{I}^{\mathrm{B}}\right) \cdot\left(\mathbb{I}^{\mathrm{V}} \otimes \tilde{\rho}^{\mathrm{AB}}\right)\right]$ \cite{cavalcanti2017all}. Eq. \eqref{rhoBMvb} describes teleportation as a collection of channels from V to B, labelled by $a$, that transform the input state $|\omega^{\text{V}}_x\rangle$ into the unnormarlized output state $\tilde{\sigma}_{a \mid \omega_x}^{\mathrm{B}}$. 
Given the structure of $M_a^\mathrm{VB}$, simple derivation shows that $M_a^\mathrm{VB}$ shares similar properties with $\frac{1}{d}(\tilde{\rho}^{\text{V}\text{B}})^{\text{T}_{\text{V}}}$, where $\tilde{\rho}^{\text{V}\text{B}}$ shares the same density matrix as $\tilde{\rho}^{\text{A}\text{B}}$ with simply replacing subsystem A by V, and  $\text{T}_{\text{A}}$ means taking partial transpose on subsystem V. Particularly, when $\{M_a^\mathrm{VA}\}_a$ is the full set of operators of standard $d\times d$-dimensional BSM (see Eq. \eqref{eq:bell_basis}), $M_a^\mathrm{VB}$ is locally equivalent with $\frac{1}{d}(\tilde{\rho}^{\text{V}\text{B}})^{\text{T}_{\text{V}}}$ in a sense that they can be transformed into each other via product local unitary transformations, and for $a=1$, $M_a^\mathrm{VB}=\frac{1}{d}(\tilde{\rho}^{\text{V}\text{B}})^{\text{T}_{\text{V}}}$ \cite{cavalcanti2017all}. Thus, combining the above discussion and Eq. \eqref{Reduction map1}, one can see that for all $a$, $M_a^\mathrm{VB}$ is constrained by
\begin{equation}
\begin{aligned}
&(\mathbb{I}^{\mathrm{V}} \otimes \Lambda^{\mathrm{B}}_{d^\prime-1}) (M_a^{\mathrm{VB}})^{\text{T}_{\text{V}}}\\
=&\text{Tr}_{\text{B}}[(M_a^{\mathrm{VB}})^{\text{T}_{\text{V}}}]\otimes \mathbb{I}^{\text{B}}-\frac{1}{d'-1} (M_a^{\mathrm{VB}})^{\text{T}_{\text{V}}} \geq 0.
\label{Reduction map2}
\end{aligned}
\end{equation}

Similarly, $N_a^{\mathrm{VB}}$ has the structure as $N_a^{\mathrm{VB}}=\operatorname{Tr}_{\mathrm{A}}\left[\left(M_a^{\mathrm{VA}} \otimes \mathbb{I}^{\mathrm{B}}\right) \cdot\left(\mathbb{I}^{\mathrm{V}} \otimes \varrho^{\mathrm{AB}}_{sep}\right)\right]$, where $\varrho^{\mathrm{AB}}_{sep}$ is an arbitrary separable state. Under this condition, $N_a^{\mathrm{VB}}$ is an unnormalized separable operator \cite{cavalcanti2017all} that satisfies $N_a^{\mathrm{VB}} \in S_1$ where $S_1$ is the set of separable operators. One can relax $S_1$ with positive partial transposition (PPT) \cite{peres1996separability}, which has a simple characterization in terms of single semidefinite constraint). 

Due to the normalization condition $\sum_a N_a^{\mathrm{VA}}=\mathbb{I}^{\mathrm{VA}}$, $\{N_a^{\mathrm{VB}}\}_a$ satisfies 
\begin{equation}
  \sum_a N_a^{\mathrm{VB}}=\mathbb{I}^{\mathrm{V}} \otimes \rho^{\mathrm{B}}_N,  
  \label{NVB}
\end{equation} 
where $\rho^{\mathrm{B}}_N$ is Bob's reduced state of $\varrho^{\mathrm{AB}}_{sep}$ that represents the generalized noise. The phrase "generalized noise" means that $\varrho^{\mathrm{AB}}_{sep}$ can be an arbitrary separable state rather than the white noise (random noise). Eq. \eqref{NVB} can also be seen as a no-signaling condition \cite{cavalcanti2017all}. To be consistent with Eq. \eqref{rhoBMvb}, $\{M_a^{\mathrm{VB}}\}_{a}$ must also satisfy 
\begin{equation}
\sum_a M_a^{\mathrm{VB}}=\mathbb{I}^{\mathrm{V}} \otimes(\sum_a \sigma_{a \mid \omega_1}^{\mathrm{B}}+\rho_N^{\mathrm{B}}).
\label{consistence condition}
\end{equation}

\begin{figure*}[htp!]
\centering
\includegraphics[width=0.8\textwidth]{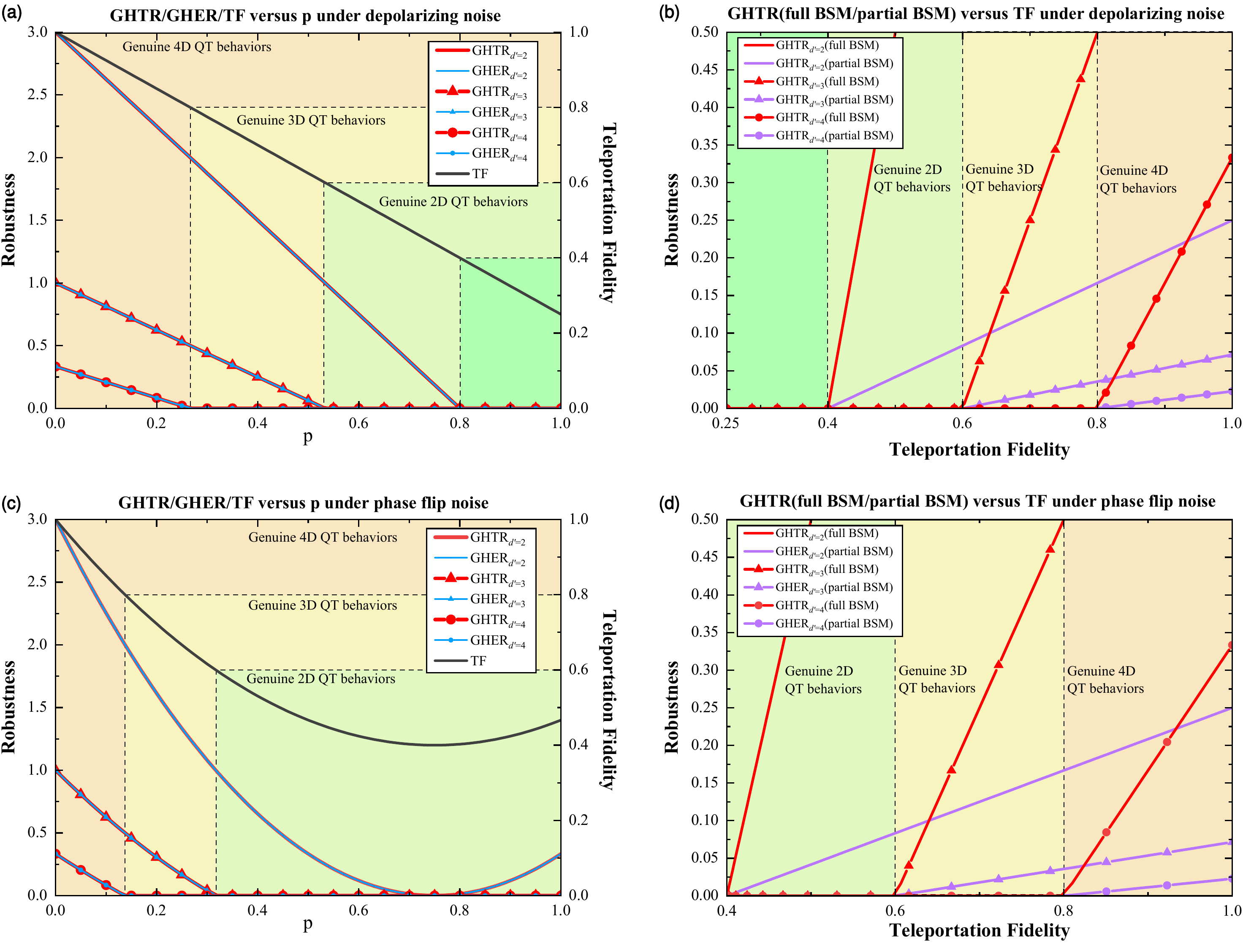}
\caption{
Numerical simulation of the genuine $d'$-dimensional teleportation robustness $\mathrm{GHTR}{d'}$, the generalized $d'$-dimensional entanglement robustness $\mathrm{GHER}{d'}$, and the average teleportation fidelity under noisy channels.
\textbf{(a, b) Depolarizing noise.} The thresholds of $\mathrm{GHTR}_{d'}$ indicate genuine 4D, 3D, and 2D QT behaviors for $0 \leq p < 0.267$, $0.267 \leq p < 0.533$, and $0.533 \leq p \leq 0.8$, respectively. Beyond $p \geq 0.8$, the process becomes simulable with separable resources. The same thresholds are obtained from $\mathrm{GHER}_{d'}$, confirming that the channel possesses 4D, 3D, and 2D entanglement in the corresponding intervals. At the transition points $p = 0.267, 0.533, 0.8$, the average teleportation fidelity equals $0.8, 0.6, 0.4$, respectively, aligning with the robustness-based criterion.
We also compute $\mathrm{GHTR}_{d'}$ using teleportation data obtained with a partial BSM (projectors ${|\phi^{\mathrm{VA}}_4\rangle\langle\phi^{\mathrm{VA}}_4|, I - |\phi^{\mathrm{VA}}_4\rangle\langle\phi^{\mathrm{VA}}_4|}$). The results scale proportionally to those obtained with a full BSM, demonstrating the feasibility of certification with incomplete measurements.
\textbf{(c, d) Phase‑flip noise.} Genuine 4D and 3D QT behaviors are certified for $0 \leq p < 0.138$ and $0.138 \leq p < 0.317$, respectively. The process remains nonclassical except at $p = 0.75$. Notably, for strong phase‑flip noise ($p > 0.75$), the teleportation fidelity exhibits a slight recovery, indicating a noise‑induced performance enhancement in this regime.
}
\label{fig: Numerical Simulation of GHTR and F}
\end{figure*}

\newcommand{\subfigref}[1]{\hyperref[fig: Numerical Simulation of GHTR and F]{\textcolor{blue}{(#1)}}}

%Consider now the case where the entanglement dimension of $\rho^{\mathrm{AB}}$ is at most $(d^\prime-1)$, i.e. the Schmidt number of $\rho^{\mathrm{AB}}$ is at most $d^\prime-1$. We denote the set of such states as $S_{d'-1}$. Usually, accurate characterization of $S_{d'-1}$ is challenging, and in many cases no analytical solution can be found. This can be remedied by relaxing the constraints slightly \cite{tavakoli2024semidefinite}. To this end, note that there exists a characterization of the set $S_{d'-1}$ in terms of positive maps \cite{terhal2000schmidt}: It holds that $\rho^{\mathrm{AB}}\in S_{d'-1}$ if and only if $(\mathbb{I}^{\mathrm{A}}\otimes \Lambda_{d'-1}^{\mathrm{B}})\rho^{\mathrm{AB}}\geq0$ for all ($d'-1$)-positive maps $\Lambda_{d'-1}^{\mathrm{B}}$. Such maps are positive when applied to one share of every bipartite state with a Schmidt number at most $(d^\prime-1)$ but non-positive for some states with a larger Schmidt number \cite{terhal2000schmidt}. This characterization is useful as it allows to define a slightly larger set than $S_{d'-1}$, which can be characterized with less effort and easy to be realized with linear programming. Hence, for all $\rho^{\mathrm{AB}}$ with Schmidt number at most $d^\prime-1$, the following constraint is satisfied,

%The key point of constructing the robustness-based criterion is to find appropriate constraints on $M_a^\mathrm{VB}$ which can be viewed as originated from state $\tilde{\rho}^\mathrm{AB}$ satisfying Eq. \eqref{Reduction map1}. 
Using Eqs. \eqref{rhoBMvb} to \eqref{consistence condition}, we construct the following convex program for certifying genuine $d^\prime$-dimensional QT behaviors of teleporting $d$-dimensional states,
\begin{small}
\begin{subequations}\label{SDP-GHTR}
\begin{align}
& \text { given }\{\sigma_{a \mid \omega_x}^{\mathrm{B}}\}_{a, x} \notag\\
& R^{d^{\prime}}_{\text{tel}}(\sigma_{a \mid \omega_x}^{\mathrm{B}})=\min _{\{N_a^{\mathrm{VB}}\},\{M_a^{\mathrm{VB}}\}} \operatorname{Tr}(\rho_N^{\mathrm{B}}) \label{SDP-GHT-1}\\
& \text { such that } \notag\\
& \sigma_{a \mid \omega_x}^{\mathrm{B}}+\operatorname{Tr}_{\mathrm{V}}\left[N_a^{\mathrm{VB}}\left(\omega_x^{\mathrm{V}} \otimes \mathbb{I}^{\mathrm{B}}\right)\right]=\operatorname{Tr}_{\mathrm{V}}\left[M_a^{\mathrm{VB}}\left(\omega_x^{\mathrm{V}} \otimes \mathbb{I}^{\mathrm{B}}\right)\right] \quad\forall x, a, \label{SDP-GHT-2}\\
& \sum_a N_a^{\mathrm{VB}}=\mathbb{I}^{\mathrm{V}} \otimes \rho_N^{\mathrm{B}}, \label{SDP-GHT-3}\\
& \sum_a M_a^{\mathrm{VB}}=\mathbb{I}^{\mathrm{V}} \otimes(\sum_a \sigma_{a \mid \omega_1}^{\mathrm{B}}+\rho_N^{\mathrm{B}}),\label{SDP-GHT-4}\\
& (\mathbb{I}^{\mathrm{V}} \otimes \Lambda^{\mathrm{B}}_{d^\prime-1}) (M_a^{\mathrm{VB}})^{\text{T}_{\text{V}}} \geq 0 \quad \forall a, \label{SDP-GHT-5}\\
& N_a^{\mathrm{VB}} \in {S_1} \quad \forall a.\label{SDP-GHT-6}
\end{align}
\end{subequations}
\end{small}
The objective of this program is to minimize the total amount of noise required to make the teleportation data simulable with lower-dimensional entanglement. The constraint \eqref{SDP-GHT-1} ensures that the moisy teleportation data (left-hand side) can be reproduced by a channel state with entanglement dimension at most $d'-1$ (right-hand side). This condition captures the idea of robustness: we ask how much noise the teleportation data can tolerate before it becomes explainable with a lower-dimensional resource. The optimal solution $r^*$ of this problem gives the minimum amount of “generalized noise” that has to be added to the teleportation data such that the mixture could be simulated by using $(d^\prime-1)$-dimensional entanglement resource. We call $R^{d^{\prime}}_{\text{tel}}(\sigma_{a \mid \omega_x}^{\mathrm{B}})=r^*$ the generalized genuine $d^{\prime}$-dimensional teleportation robustness ($\text{GHTR}_{d^\prime}$) of the data $\{\sigma_{a \mid \omega_x}^{\mathrm{B}}\}_{a, x}$. Hence the robustness-based criterion is
\begin{equation}
    R^{d^{\prime}}_{\text{tel}}(\sigma_{a \mid \omega_x}^{\mathrm{B}})>0.
\label{Robustness-based criterion}
\end{equation}
Satisfying Eq. \eqref{Robustness-based criterion} demonstrates at least genuine $d^\prime$-dimensional QT behavior in teleporting $d$-dimensional states. Note that since the constraints on variables $M_a^{\mathrm{VB}}$ are obtained from employing an outer relaxation (Eq. \eqref{Reduction map}) on set $S_{d'-1}$, hence the robustness-based criterion \eqref{Robustness-based criterion} detects genuine $d'$-dimensional QT behaviors that are originated from a slightly smaller set of states with $d'$-dimensional entanglement. If an accurate characterization of set $S_{d'-1}$ exists, the criterion can detect all $d'$-dimensional QT behaviors, as we theoretically prove in Sec. \ref{4:B} and App. \ref{Ap:F}. Though such characterization is yet unknown, criterion \eqref{Robustness-based criterion} based on relaxation \eqref{Reduction map} still has stronger detection ability than the fidelity-based criterion \eqref{result 1}, as we show in Sec. \ref{4:C}.

\subsection{\label{4:B}Application: demonstrating that all channel states with $d'$-dimensional entanglement enable genuine $d'$-dimensional QT behavior}

Whether quantum entanglement and quantum teleportation capacity can be viewed as equivalent is a fundamental question in quantum information. However, there exist bound entangled states whose teleportation fidelity could not exceed the classical limit \cite{horodecki1999general}, leading to a long-held view that a mismatch lies between the two concepts. In 2017, it was proven that all entangled states enable nonclassical teleportation with a new metric-teleportation robustness \cite{cavalcanti2017all}, eliminating the mismatch and consequently solving the question in nonclassical QT aspect. However this question in genuine HDQT aspect has not been answered, i.e. whether $d'$-dimensional quantum entanglement and genuine $d'$-dimensional quantum teleportation capacity can be viewed as equivalent, which is critical to the accurate evaluation of HD entangled resource. 

To fill this gap, we prove that (see Appendix ~\ref{Ap:F} for details), in the case that (i) at least one of Alice's measurement operators corresponds to a projection onto a maximally entangled state and (ii) the inputs $|\omega^{\mathrm{V}}_x\rangle$ are tomographically complete, $R^{d^{\prime}}_{\text{tel}}(\sigma_{a \mid \omega_x}^{\mathrm{B}})$ is proportional to $R^{d^{\prime}}_{\text{ent}}(\rho^{\mathrm{AB}})$, the generalized $d'$-dimensional entanglement robustness ($\mathrm{GHER}_{d'}$) of the state $\rho^{\mathrm{AB}}$,
\begin{subequations}
\begin{align}
& \text { given } \rho^{\mathrm{AB}} \notag\\
& R_{\text {ent }}^{d^{\prime}}(\rho^{\mathrm{AB}})=\min _{N^{\mathrm{AB}}, M^{\mathrm{AB}}} \text{Tr}(N^{\mathrm{AB}}) \label{SDP ENT-1}\\
& \text { such that } \notag\\
& \rho^{\mathrm{AB}}+ N^{\mathrm{AB}}=M^{\mathrm{AB}}, \label{SDP ENT-2}\\
& (\mathbb{I}^{\mathrm{V}} \otimes \Lambda^{\mathrm{B}}_{d^\prime-1}) M^{\mathrm{AB}} \geq 0,\label{SDP ENT-3}\\
& N^{\mathrm{AB}} \in S, \label{SDP ENT-4}
\end{align}
\label{SDP ENT}
\end{subequations}
where $N^{\mathrm{AB}}$ is a separable Hermitian operator that represents the generalized noise. This SDP is an variant of the one in Ref. \cite{cobucci2024detecting} by choosing the generalized noise instead of the random noise. Moreover, if Alice performs full BSM, then $R^{d^{\prime}}_{\text{tel}}(\sigma_{a \mid \omega_x}^{\mathrm{B}})=R^{d^{\prime}}_{\text{ent}}(\rho^{\mathrm{AB}})$. Since $R^{d^{\prime}}_{\text{ent}}(\rho^{\mathrm{AB}})$ is non-null if and only if $\rho^{\mathrm{AB}}$ has Schmidt number at least $d'$, all $d'$-dimensional entangled states that have positive $\text{GHER}_{d'}$ can demonstrate genuine $d'$-dimensional QT. Notably, if the characterization of $S_{d'-1}$ is accurate, rather than its outer relaxation, all $d'$-dimensional entangled states can demonstrate genuine $d'$-dimensional QT.

Here we discuss the validity of above conclusion through the numerical simulation of two concrete noisy HDQT examples. We consider the four-dimensional teleportation scenario where the input $\{\omega^{\mathrm{V}}_x\}$ is the complete set of states from four-dimensional mutually unbiased bases (MUBs \cite{durt2010mutually}).
To model realistic experimental imperfections, we examine two representative types of noise that commonly affect HD photonic or atomic systems: depolarizing noise and phase-flip noise. These models capture distinct physical mechanisms of decoherence, allowing us to test the robustness of our certification criteria under different error sources.

\emph{Depolarizing noise (dp)}-This model describes a situation where the quantum system interacts isotropically with its environment, causing the state to become completely mixed with probability $p$. Such noise may arise from uncontrolled interactions, thermal fluctuations, or imperfect state preparation. The corresponding channel state is:
\begin{equation}
    \rho_1^{\mathrm{AB}}=(1-p)|\phi^{\mathrm{AB}}_4\rangle\langle\phi^{\mathrm{AB}}_4|+p\frac{\mathbb{I}^{\mathrm{AB}}}{16},
\end{equation}
where $|\phi^{\mathrm{AB}}_4\rangle$ is the maximally four-dimensional entangled state $|\phi^{\mathrm{AB}}_4\rangle=\sum_{i=0}^{3}(1/2)|i\rangle|i\rangle$. 

\emph{Phase‑flip noise (pf)}-This noise model specifically affects the relative phases between computational basis states, which is a dominant error mechanism in many HD encodings (e.g., orbital angular momentum or time‑bin photonic qudits). Phase flips can originate from path‑length fluctuations, ambient magnetic fields, or imperfect phase stabilization. The corresponding channel state is:
\begin{equation}
\rho_2^{\mathrm{AB}}=\sum^{4}_{i,j=1}(E^{pf}_i \otimes E^{pf}_j)|\phi^{\mathrm{AB}}_4\rangle\langle\phi^{\mathrm{AB}}_4|(E^{pf}_i \otimes E^{pf}_j)^\dagger,
\end{equation}
where $E^{pf}_1=\sqrt{1-p}\tilde{U}_{00},E^{pf}_2=\sqrt{p/3}\tilde{U}_{10},E^{pf}_3=\sqrt{p/3}\tilde{U}_{20},E^{pf}_4=\sqrt{p/3}\tilde{U}_{30}$ (the superscript (pf) denotes phase flip noise) are the corresponding  Kraus operators ($\tilde{U}_{nm}=\sum^{4}_{k=1}e^{\pi \mathrm{i}kn/2}|k\rangle\langle k\bigoplus m| $) \cite{fonseca2019high}. 

The Numerical results of program \eqref{SDP-GHTR}, program \eqref{SDP ENT} and average teleportation fidelity are presented in Fig. \ref{fig: Numerical Simulation of GHTR and F} (The details of the simulations can be found in Ref. \cite{gong_2026_18366412}). As shown in \ref{fig: Numerical Simulation of GHTR and F}\subfigref{a}, under depolarizing noise, $\mathrm{GHTR}_{d'}$ indicates genuine 4D, 3D, and 2D QT behaviors for $0 \leq p < 0.267$, $0.267 \leq p < 0.533$, and $0.533 \leq p \leq 0.8$, respectively. When $p \geq 0.8$, the teleportation process is degraded into classical region, i.e., simulable with separable resources. The same thresholds are obtained from $\mathrm{GHER}_{d'}$, confirming that the channel possesses 4D, 3D, and 2D entanglement in the corresponding intervals. At the transition points $p = 0.267, 0.533, 0.8$, the average teleportation fidelity equals $0.8, 0.6, 0.4$, respectively, aligning with the robustness-based criterion. As shown in \ref{fig: Numerical Simulation of GHTR and F}\subfigref{c}, under phase flip noise, genuine 4D and 3D QT behaviors are certified for $0 \leq p < 0.138$ and $0.138 \leq p < 0.317$, respectively. The process remains nonclassical except at $p = 0.75$. Notably, for strong phase‑flip noise ($p > 0.75$), the teleportation fidelity exhibits a slight recovery, indicating a noise‑induced performance enhancement in this regime.

The numerical results confirm the theoretical link between teleportation and entanglement robustness. For full Bell‑state measurements (Fig. \ref{fig: Numerical Simulation of GHTR and F}\subfigref{a} and \subfigref{b}), $\text{GHTR}_{d'}$ is equal to $\text{GHER}_{d'}$, demonstrating that the teleportation data directly reflect the dimensionality of the channel entanglement. When only a partial BSM is employed (Fig. \ref{fig: Numerical Simulation of GHTR and F}\subfigref{b} and \subfigref{d}), the teleportation robustness remains proportional to the entanglement robustness, underscoring the practical feasibility of certification even with incomplete measurements. All these results align with and validate the theoretical statements derived in Sec. \ref{sec:4:A} and Appendix \ref{Ap:F}.

\subsection{\label{4:C}Application: identifying "weak" HDQT behaviors}

Here we compare the certification ability of the robustness-based criterion \eqref{Robustness-based criterion} with the fidelity-based criterion \eqref{result 1} and show that the robustness-based criterion has stronger noise resilience, as it can identify those "weak" HDQT behaviors that the fidelity-based criterion fails to detect.

Consider now that in the case of teleporting $d$-dimensional states, the maximally entangled channel state $|\phi^{\mathrm{AB}}_d\rangle$ is affected by the amplitude damping (ad) noise whose Kraus operators are $E^{ad}_1=|0\rangle\langle 0|+\sqrt{1-p} \sum_{j=1}^{d-1}|j\rangle\langle j|$ and $E^{ad}_j=\sqrt{p}|0\rangle\langle j-1|$ with $j=2, \cdots, d$ ($0\leq p \leq1$) \cite{fonseca2019high} (the superscript (ad) denotes amplitude damping noise). We focus on $d=3,4$ with Alice performing full BSM, and numerically compute the maximal value of noise parameter (denoted by $p_{\mathrm{mv}}$) that the two criteria can tolerate such that the genuine $d'$-dimensional QT behavior can be certified. 

As shown in Table \ref{tab:maintext-table1}, compared with the fidelity-based criterion, the noise strength that the robustness-based criterion can tolerate is higher for all cases considered. This indicates that the robustness-based criterion exhibits stronger noise resistance. Physically, this is because the robustness metric directly quantifies how much noise must be added to the teleportation data before it becomes simulable with lower-dimensional entanglement, whereas the fidelity criterion depends on a single averaged quantity that can be more sensitive to specific types of noise. 

Noteworthy, the results reveal that, similar to bound entanglement (BE) which cannot be detected by the fidelity-based nonclassical teleportation  criteria \cite{horodecki1999general}, $\bar{F}_{\mathrm{tel}}^{B E}<2 /(d+1)$, there also exists bound $d'$-dimensional entanglement (BHDE)  that leads to  $\bar{F}_{\text {tel }}^{B H D E}<d^{\prime} /(d+1)$. This phenomenon highlights the subtle dissimilarity between HDQT fidelity and HD entanglement. Notably, such a mismatching can be corrected by the utilization of robustness-based criterion since $\mathrm{GHTR}_{d'}$, which turn out to be identical, as shown in Table \ref{tab:maintext-table1} and also proved in Appendix ~\ref{Ap:F}, to $\mathrm{GHER}_{d'}$.

\begin{table}
\caption{\label{tab:maintext-table1}Maximum tolerable noise strength $p_{\mathrm{mv}}$ for certifying genuine $d'$-dimensional teleportation via the fidelity-based criterion $p_{\mathrm{mv}}(\bar{F}_{\mathrm{tel}})$ and the robustness-based criterion $p_{\mathrm{mv}}(\mathrm{GHTR}_{d'})$. 
The column $p_{\mathrm{mv}}(\mathrm{GHER}_{d'})$ shows the corresponding threshold for certifying $d'$-dimensional entanglement of the channel state. 
Under full Bell‑state measurements, the theoretical equality $\mathrm{GHTR}_{d'} = \mathrm{GHER}_{d'}$ (derived in Sec. \ref{sec:4:A} and Appendix \ref{Ap:F}) implies identical noise tolerances, as confirmed numerically here. The table highlights the consistently higher noise resilience of the robustness criterion compared to the fidelity benchmark.}
\begin{ruledtabular}
\begin{tabular}{cccc}
 $(d,d')$ &$p_{\mathrm{mv}}(\bar{F}_{\mathrm{tel}})$ &$p_{\mathrm{mv}}(\mathrm{GHTR}_{d'})$ &$p_{\mathrm{mv}}(\mathrm{GHER}_{d'})$\\
\hline
$(3,3)$& 0.292 & 0.323 & 0.323\\
$(4,3)$& 0.422 & 0.492 & 0.492\\
$(4,4)$& 0.183 & 0.21 & 0.21\\
\end{tabular}
\end{ruledtabular}
\end{table}

\section{\label{5}Summary and outlook}

In this work, we have established a comprehensive theoretical framework for certifying genuine HDQT behaviors. We introduced two universal criteria that rigorously rule out simulations using lower-dimensional entanglement, solely through the analysis of input-output teleportation data.

Our criteria offer a significant advance beyond previous methods \cite{hu2020experimental,luo2019quantum}, which could certify the transmission of a genuine high-dimensional state but not the entanglement dimension of the resource. This distinction is critical: the latter guarantees the teleportation channel’s full capacity and noise resilience, which are essential for reliable HD quantum networks.

The immediate experimental relevance of our framework lies in its relaxed requirements and direct feasibility on near-term platforms. Both criteria are provably effective with partial Bell-state measurements (BSM), a feature that dramatically reduces experimental overhead. In photonic systems, for instance, performing a complete, $d^2$-outcome BSM is a formidable technical challenge, often requiring complex multi-port interferometers or ancillary photons. With our criteria, such full BSM is unnecessary for certification. Experimentalists can instead implement a significantly simpler measurement, such as a binary “success/failure” projection onto a single Bell state. This significantly reduces the number of required measurement bases, thereby transforming the experimental complexity from often prohibitive to readily manageable.

Importantly, techniques for efficient HD state tomography are well-developed \cite{pereira2018adaptive,rambach2021robust,zhou2021direct1}. Our criteria do not demand new, unproven technology; they provide a rigorous and optimized analysis protocol to be applied to data that can be collected with existing tools. The numerical thresholds we provide serve as quantitative targets for experimentalists to benchmark their systems against.

Looking forward, this work sets a new, indispensable standard for validating the core quantum advantage in HDQT. As the field progresses towards more complex high-dimensional quantum networks and repeaters, verifying the true dimensionality of the teleportation channel will be paramount. Our universally applicable criteria—feasible under realistic experimental constraints—provide the necessary tool for this task. They not only guide the design of more robust and efficient teleportation experiments across all physical platforms but also establish a common benchmark for assessing and comparing the performance of high-dimensional quantum links, thereby accelerating the development of a reliable high-dimensional quantum internet.

\begin{acknowledgments}

This work is supported in part by the National Natural Science Foundation of China under Grant No. 62471058, and in part by the Fund of State Key Laboratory of Information Photonics and Optical Communications, Beijing
University of Posts and Telecommunications, China (No.IPOC2022ZT10).
%\dots.
\end{acknowledgments}

\bibliography{reference}

\onecolumngrid
\appendix

\section{\label{Ap:A}Demonstrating the transmission of genuine high-dimensional state is insufficient for verifying genuine HDQT}

In this section, we charify the necessity of constructing the new criteria for the dimension certification of HDQT, as previous methods \cite{luo2019quantum,hu2020experimental}, although effective in checking the dimension of output states, are not sufficient for certifying the existence of HD entanglement.

In previous work, the verification of genuine HDQT focus on demonstrating that the experimental setup possess the ability of teleporting a genuine qutrit that cannot be described by an incoherent mixture of qubits. Based on this point, the authors in Ref. \cite{luo2019quantum} provide a genuine three-dimensional QT criterion $F^{\psi_0}_{\text {tel }}>\frac{2}{3}$ where $F^{\psi_0}_{\text {tel }}$ is the measured teleportation fidelity when the input state is $|\psi^{\mathrm{V}}_0\rangle=(|0^{\mathrm{V}}\rangle+|1^{\mathrm{V}}\rangle+|2^{\mathrm{V}}\rangle)/\sqrt{3}$. And Ref. \cite{hu2020experimental} followed the same idea and proposed a robustness criterion $\mu>0$ where $\mu$ is the minimum amount of “white noise” that must be added to the qutrit state such that the mixture can be simulated by qubit states. It was shown that this robustness criterion is useful for a broader scope of input states with the form of $|\psi^{\mathrm{V}}\rangle=(|0^{\mathrm{V}}\rangle+e^{\mathrm{i}\theta_1}|1^{\mathrm{V}}\rangle+e^{\mathrm{i}\theta_2}|2^{\mathrm{V}}\rangle)/\sqrt{3}$ ($\theta_1,\theta_2$ refer to the phase information and $\theta_1,\theta_2 \in[0,2\pi)$). 

These two criteria are useful for experimenters to test whether the setup possesses the capability of teleporting genuine three-dimensional states. However, the verification of genuine HDQT must also certify whether the dimension of entanglement meet the necessary threshold, which directly determines the transmission capacity and noise resilience. In the following, we show that the these two criteria are not sufficient for certifying the existence of HD entanglement by showing that there exist a simulation strategy that can pass both of the criteria with local operations and 
 only lower-dimensional entanglement.

The detailed simulation strategy can be presented as follows. After receiving the input state $|\psi\rangle$, the device of the sender executes a local three-dimensional discrete Fourier transform $\mathcal{QFT}_3$ \cite{weinstein2001implementation} on the input state and obtains
\begin{equation}
|\psi'^{\mathrm{V}}\rangle=\mathcal{QFT}_3 |\psi^{\mathrm{V}}\rangle=\frac{1}{\sqrt{3}}\left(\begin{array}{ccc}
1 & 1 & 1 \\
1 & e^{\frac{2}{3} \pi \mathrm{i}} & e^{\frac{4}{3} \pi \mathrm{i}} \\
1 & e^{\frac{4}{3} \pi \mathrm{i}} & e^{\frac{2}{3} \pi \mathrm{i}}
\end{array}\right) \frac{1}{\sqrt{3}}\left(\begin{array}{l}
1 \\
e^{\mathrm{i}\theta_1} \\
e^{\mathrm{i}\theta_2}
\end{array}\right)=\frac{1}{3}
\left(\begin{array}{l}
1+e^{\mathrm{i}\theta_1}+e^{\mathrm{i}\theta_2} \\
1+e^{\mathrm{i}(\frac{2\pi}{3}+
\theta_1)}+e^{\mathrm{i}(\frac{4\pi}{3}+\theta_2)} \\
1+e^{\mathrm{i}(\frac{4\pi}{3}+
\theta_1)}+e^{\mathrm{i}(\frac{2\pi}{3}+\theta_2)}
\end{array}\right).
\end{equation}
This local preprocessing step is crucial: it effectively "encodes" the three-dimensional input state into a two-dimensional subspace that can be transmitted using a two-dimensional Bell state with better performance. Then, $|\psi'^{\mathrm{V}}\rangle$ is teleported by a three-dimensional quantum teleportation scheme with two-dimensional Bell state  $|\varphi^{\mathrm{AB}}\rangle=\frac{1}{\sqrt{2}}(|0^{\mathrm{A}}\rangle|0^{\mathrm{B}}\rangle+|1^{\mathrm{A}}\rangle|1^{\mathrm{B}}\rangle)$ as channel state. The scheme is pesented as follows. 

\emph{Scheme of teleporting three-dimensional states with two-dimensional Bell state.} Here we introduce the scheme in a general manner such that the input state is an arbitrary three-dimensional pure state, which is denoted as  $|\omega^\mathrm{V}\rangle=\alpha|0\rangle+\beta|1\rangle+\gamma|2\rangle$ satisfying the normalization condition $\left|\alpha\right|^2+\left|\beta\right|^2+\left|\gamma\right|^2=1$. The channel state used is $|\varphi^{\mathrm{AB}}\rangle=\frac{1}{\sqrt{2}}(|0^{\mathrm{A}}\rangle|0^{\mathrm{B}}\rangle+|1^{\mathrm{A}}\rangle|1^{\mathrm{B}}\rangle)$. Hence the combined state of systems V, A and B is 
\begin{equation}
|\Psi^{\mathrm{VAB}}\rangle=\frac{1}{\sqrt{2}}\left(\alpha|0^{\mathrm{V}}\rangle|0^{\mathrm{A}}\rangle|0^{\mathrm{B}}\rangle+\alpha|0^{\mathrm{V}}\rangle|1^{\mathrm{A}}\rangle|1^{\mathrm{B}}\rangle+\beta|1^{\mathrm{V}}\rangle|0^{\mathrm{A}}\rangle|0^{\mathrm{B}}\rangle+\beta|1^{\mathrm{V}}\rangle|1^{\mathrm{A}}\rangle|1^{\mathrm{B}}\rangle+\gamma|2^{\mathrm{V}}\rangle|0^{\mathrm{A}}\rangle|0^{\mathrm{B}}\rangle+\gamma|2^{\mathrm{V}}\rangle|1^{\mathrm{A}}\rangle|1^{\mathrm{B}}\rangle\right).
\label{system VAB}
\end{equation}

The measurement performed by sender Alice is the standard BSM, whose measurement basis is the set of $3 \times 3$ orthogonal Bell states given by $\left|\phi_{m n}\right\rangle=\frac{1}{\sqrt{3}} \sum_{l=0}^{2} e^{\frac{2 \pi \mathrm{i}}{3} l m }|l\rangle \otimes|l+n\rangle$. Here, $n=0,1,2$ (where $l+n$ must be taken from modulo $3$) denotes the trit information of the three-dimensional state and $m=0,1,2$ represents the relative phase information. The explicit form of $\{\left|\phi_{m n}\right\rangle\}$ is
\begin{equation}
\begin{aligned}
&\left|\phi_{00}\right\rangle=\frac{1}{\sqrt{3}}(|00\rangle+|11\rangle+|22\rangle),&\left|\phi_{10}\right\rangle=\frac{1}{\sqrt{3}}\left(|00\rangle+e^{\frac{2 \pi \mathrm{i}}{3}}|11\rangle+e^{\frac{4 \pi \mathrm{i}}{3}}|22\rangle\right),&\left|\phi_{20}\right\rangle=\frac{1}{\sqrt{3}}\left(|00\rangle+e^{\frac{4 \pi \mathrm{i}}{3}}|11\rangle+e^{\frac{2 \pi \mathrm{i}}{3}}|22\rangle\right), \\
&\left|\phi_{01}\right\rangle=\frac{1}{\sqrt{3}}(|01\rangle+|12\rangle+|20\rangle),&\left|\phi_{11}\right\rangle=\frac{1}{\sqrt{3}}\left(|01\rangle+e^{\frac{2 \pi \mathrm{i}}{3}}|12\rangle+e^{\frac{4 \pi \mathrm{i}}{4}}|20\rangle\right),&\left|\phi_{21}\right\rangle=\frac{1}{\sqrt{3}}\left(|01\rangle+e^{\frac{4 \pi \mathrm{i}}{3}}|12\rangle+e^{\frac{2 \pi \mathrm{i}}{3}}|20\rangle\right), \\
&\left|\phi_{02}\right\rangle=\frac{1}{\sqrt{3}}(|02\rangle+|10\rangle+|21\rangle),&\left|\phi_{12}\right\rangle=\frac{1}{\sqrt{3}}\left(|02\rangle+e^{\frac{2 \pi \mathrm{i}}{3}}|10\rangle+e^{\frac{4 \pi \mathrm{i}}{3}}|21\rangle\right),&\left|\phi_{22}\right\rangle=\frac{1}{\sqrt{3}}\left(|02\rangle+e^{\frac{4 \pi \mathrm{i}}{3}}|10\rangle+e^{\frac{2 \pi \mathrm{i}}{3}}|21\rangle\right).
\end{aligned}
\label{3dBSM}
\end{equation}

\begin{table}
\caption{\label{tab:table1}The full information about the unitary operations, the output states, the fidelity functions and the projection probabilities for all projection results $\left|\phi_{mn}^{\mathrm{VA}}\right\rangle$ of BSM.}
\begin{ruledtabular}
\begin{tabular}{ccccc}
 projection result &unitary operation $U_{mn}$ &output state $|\Psi^{\mathrm{B}}_{mn}\rangle$ &fidelity function $F_{mn}$&
 projection probability $p_{mn}$ \\
\hline
$|\phi^{\mathrm{VA}}_{00}\rangle$& $\left(\begin{array}{lll}
1 & 0 & 0 \\
0 & 1 & 0 \\
0 & 0 & 1
\end{array}\right)$ & $\frac{\alpha|0^{\mathrm{B}}\rangle+\beta|1^{\mathrm{B}}\rangle}{\sqrt{|\alpha|^2+|\beta|^2}}$ & $|\alpha|^2+|\beta|^2$ &$\frac{|\alpha|^2+|\beta|^2}{6}$\\
$|\phi^{\mathrm{VA}}_{10}\rangle$& $\left(\begin{array}{ccc}
1 & 0 & 0 \\
0 & e^{\frac{2 \pi \mathrm{i}}{3}} & 0 \\
0 & 0 & 1
\end{array}\right)$ & $\frac{\alpha|0^{\mathrm{B}}\rangle+\beta|1^{\mathrm{B}}\rangle}{\sqrt{|\alpha|^2+|\beta|^2}}$ & $|\alpha|^2+|\beta|^2$ &$\frac{|\alpha|^2+|\beta|^2}{6}$\\
$|\phi^{\mathrm{VA}}_{20}\rangle$& $\left(\begin{array}{ccc}
1 & 0 & 0 \\
0 & e^{\frac{4 \pi \mathrm{i}}{3}} & 0 \\
0 & 0 & 1
\end{array}\right)$ & $\frac{\alpha|0^{\mathrm{B}}\rangle+\beta|1^{\mathrm{B}}\rangle}{\sqrt{|\alpha|^2+|\beta|^2}}$ & $|\alpha|^2+|\beta|^2$ &$\frac{|\alpha|^2+|\beta|^2}{6}$\\
$|\phi^{\mathrm{VA}}_{01}\rangle$& $\left(\begin{array}{ccc}
0 & 1 & 0 \\
0 & 0 & 1 \\
1 & 0 & 0
\end{array}\right)$ & $\frac{\alpha|0^{\mathrm{B}}\rangle+\gamma|2^{\mathrm{B}}\rangle}{\sqrt{|\alpha|^2+|\gamma|^2}}$ & $|\alpha|^2+|\gamma|^2$ &$\frac{|\alpha|^2+|\gamma|^2}{6}$\\
$|\phi^{\mathrm{VA}}_{11}\rangle$& $\left(\begin{array}{ccc}
0 & 1 & 0 \\
0 & 0 & 1 \\
e^{\frac{4 \pi \mathrm{i}}{3}} & 0 & 0
\end{array}\right)$ & $\frac{\alpha|0^{\mathrm{B}}\rangle+\gamma|2^{\mathrm{B}}\rangle}{\sqrt{|\alpha|^2+|\gamma|^2}}$ & $|\alpha|^2+|\gamma|^2$ &$\frac{|\alpha|^2+|\gamma|^2}{6}$\\
$|\phi^{\mathrm{VA}}_{21}\rangle$& $\left(\begin{array}{ccc}
0 & 1 & 0 \\
0 & 0 & 1 \\
e^{\frac{2 \pi \mathrm{i}}{3}} & 0 & 0
\end{array}\right)$ & $\frac{\alpha|0^{\mathrm{B}}\rangle+\gamma|2^{\mathrm{B}}\rangle}{\sqrt{|\alpha|^2+|\gamma|^2}}$ & $|\alpha|^2+|\gamma|^2$ &$\frac{|\alpha|^2+|\gamma|^2}{6}$\\
$|\phi^{\mathrm{VA}}_{02}\rangle$& $\left(\begin{array}{ccc}
0 & 0 & 1 \\
1 & 0 & 0 \\
0 & 1 & 0
\end{array}\right)$ & $\frac{\beta|1^{\mathrm{B}}\rangle+\gamma|2^{\mathrm{B}}\rangle}{\sqrt{|\beta|^2+|\gamma|^2}}$ & $|\beta|^2+|\gamma|^2$ &$\frac{|\beta|^2+|\gamma|^2}{6}$\\
$|\phi^{\mathrm{VA}}_{12}\rangle$& $\left(\begin{array}{ccc}
0 & 0 & 1 \\
e^{\frac{2 \pi \mathrm{i}}{3}} & 0 & 0 \\
0 & e^{\frac{4 \pi \mathrm{i}}{3}} & 0
\end{array}\right)
$ & $\frac{\beta|1^{\mathrm{B}}\rangle+\gamma|2^{\mathrm{B}}\rangle}{\sqrt{|\beta|^2+|\gamma|^2}}$ & $|\beta|^2+|\gamma|^2$ &$\frac{|\beta|^2+|\gamma|^2}{6}$\\
$|\phi^{\mathrm{VA}}_{22}\rangle$& $\left(\begin{array}{ccc}
0 & 0 & 1 \\
e^{\frac{4 \pi \mathrm{i}}{3}} & 0 & 0 \\
0 & e^{\frac{2 \pi \mathrm{i}}{3}} & 0
\end{array}\right)$ & $\frac{\beta|1^{\mathrm{B}}\rangle+\gamma|2^{\mathrm{B}}\rangle}{\sqrt{|\beta|^2+|\gamma|^2}}$ & $|\beta|^2+|\gamma|^2$ &$\frac{|\beta|^2+|\gamma|^2}{6}$\\
\end{tabular}
\end{ruledtabular}
\end{table}

Eq. \eqref{system VAB} can be rewritten in the basis of Eq. \eqref{3dBSM} as 
\begin{equation}
\begin{aligned}
|\Psi^{\mathrm{VAB}}\rangle =\frac{1}{\sqrt{6}}&\left[\alpha\left(|\phi^{\mathrm{VA}}_{00}\rangle+|\phi^{\mathrm{VA}}_{10}\rangle+|\phi^{\mathrm{VA}}_{20}\rangle\right)|0^{\mathrm{B}}\rangle+\alpha\left(|\phi^{\mathrm{VA}}_{01}\rangle+|\phi^{\mathrm{VA}}_{11}\rangle+|\phi^{\mathrm{VA}}_{21}\rangle\right)|1^{\mathrm{B}}\rangle\right. \\ +&\beta\left(|\phi^{\mathrm{VA}}_{02}\rangle+e^{\frac{4 \pi \mathrm{i}}{3}}|\phi^{\mathrm{VA}}_{12}\rangle+e^{\frac{2 \pi \mathrm{i}}{3}}|\phi^{\mathrm{VA}}_{22}\rangle\right)|0^{\mathrm{B}}\rangle+\beta\left(|\phi^{\mathrm{VA}}_{00}\rangle+e^{\frac{4 \pi \mathrm{i}}{3}}|\phi^{\mathrm{VA}}_{10}\rangle+e^{\frac{2 \pi \mathrm{i}}{3}}|\phi^{\mathrm{VA}}_{20}\rangle\right)|1^{\mathrm{B}}\rangle \\ +&\left.\gamma\left(|\phi^{\mathrm{VA}}_{01}\rangle+e^{\frac{2 \pi \mathrm{i}}{3}}|\phi^{\mathrm{VA}}_{11}\rangle+e^{\frac{4 \pi \mathrm{i}}{3}}|\phi^{\mathrm{VA}}_{21}\rangle\right)|0^{\mathrm{B}}\rangle+\gamma\left(|\phi^{\mathrm{VA}}_{02}\rangle+e^{\frac{2 \pi \mathrm{i}}{3}}|\phi^{\mathrm{VA}}_{12}\rangle+e^{\frac{4 \pi \mathrm{i}}{3}}|\phi^{\mathrm{VA}}_{22}\rangle\right)|1^{\mathrm{B}}\rangle\right] \\
=\frac{1}{\sqrt{6}}&\left[|\phi^{\mathrm{VA}}_{00}\rangle(\alpha|0^{\mathrm{B}}\rangle+\beta|1^{\mathrm{B}}\rangle)+|\phi^{\mathrm{VA}}_{10}\rangle(\alpha|0^{\mathrm{B}}\rangle+e^{\frac{4 \pi \mathrm{i}}{3}}\beta|1^{\mathrm{B}}\rangle)+|\phi^{\mathrm{VA}}_{20}\rangle(\alpha|0^{\mathrm{B}}\rangle+e^{\frac{2 \pi \mathrm{i}}{3}}\beta|1^{\mathrm{B}}\rangle)\right.\\
+&|\phi^{\mathrm{VA}}_{01}\rangle(\gamma|0^{\mathrm{B}}\rangle+\alpha|1^{\mathrm{B}}\rangle)+|\phi^{\mathrm{VA}}_{11}\rangle(\gamma e^{\frac{2 \pi \mathrm{i}}{3}}|0^{\mathrm{B}}\rangle+\alpha|1^{\mathrm{B}}\rangle)+|\phi^{\mathrm{VA}}_{21}\rangle(\gamma e^{\frac{4 \pi \mathrm{i}}{3}}|0^{\mathrm{B}}\rangle+\alpha|1^{\mathrm{B}}\rangle)\\
+&\left.|\phi^{\mathrm{VA}}_{02}\rangle(\beta|0^{\mathrm{B}}\rangle+\gamma|1^{\mathrm{B}}\rangle)+|\phi^{\mathrm{VA}}_{12}\rangle(\beta e^{\frac{4 \pi \mathrm{i}}{3}}|0^{\mathrm{B}}\rangle+\gamma e^{\frac{2 \pi \mathrm{i}}{3}}|1^{\mathrm{B}}\rangle)+|\phi^{\mathrm{VA}}_{22}\rangle(\beta e^{\frac{2 \pi \mathrm{i}}{3}}|0^{\mathrm{B}}\rangle+\gamma  e^{\frac{4 \pi \mathrm{i}}{3}}|1^{\mathrm{B}}\rangle)\right].
\end{aligned}
\label{VAB1}
\end{equation}

For the case that systems V and A are projected onto $|\phi^{\mathrm{VA}}_{00}\rangle$, the unnormalized output state is given by
\begin{equation}
|\widetilde{\Psi}_{00}^\mathrm{B}\rangle=\frac{1}{\sqrt{6}}(\alpha|0^{\mathrm{B}}\rangle+\beta|1^{\mathrm{B}}\rangle)=\sqrt{p_{00}}\left|\Psi_{00}^\mathrm{B}\right\rangle
=\frac{\sqrt{|\alpha|^2+|\beta|^2}}{\sqrt{6}}\left(\frac{\alpha}{\sqrt{|\alpha|^2+|\beta|^2}}|0^{\mathrm{B}}\rangle+\frac{\beta}{\sqrt{|\alpha|^2+|\beta|^2}}|1^{\mathrm{B}}\rangle\right),
\end{equation}
where $p_{00}=\frac{|\alpha|^2+|\beta|^2}{6}$ denotes the probability of the projection onto $|\phi^{\mathrm{VA}}_{00}\rangle$. Then receiver Bob performs unitary operation $I=\mathrm{Diag}\{1,1,1\}$ to obtain the teleportation fidelity as
\begin{equation}
F_{00}=\langle \omega|\Psi_{00}\rangle\langle \Psi_{00}|\omega\rangle=\frac{1}{|\alpha|^2+|\beta|^2}\left(\begin{array}{lll}
\alpha^* & \beta^* & \gamma^*
\end{array}\right)\left(\begin{array}{ccc}
|\alpha|^2 & \alpha \beta^* & 0 \\
\beta \alpha^* & |\beta|^2 & 0 \\
0 & 0 & 0
\end{array}\right)\left(\begin{array}{l}
\alpha \\
\beta \\
\gamma
\end{array}\right)=|\alpha|^2+|\beta|^2.
\end{equation}

For all cases where systems V and A are projected onto $|\phi^{\mathrm{VA}}_{mn}\rangle$, the corresponding unitary operations, the output states, the fidelity functions and the projection probabilities are concluded in Table \ref{tab:table1}. Noteworthy, those unitary operations are only determined by the channel resource and the outcomes of Alice, not conflicting with the assumptions regarding the transmission of unknown states.

\begin{figure*}[htp!]
\centering
\includegraphics[width=0.5\textwidth]{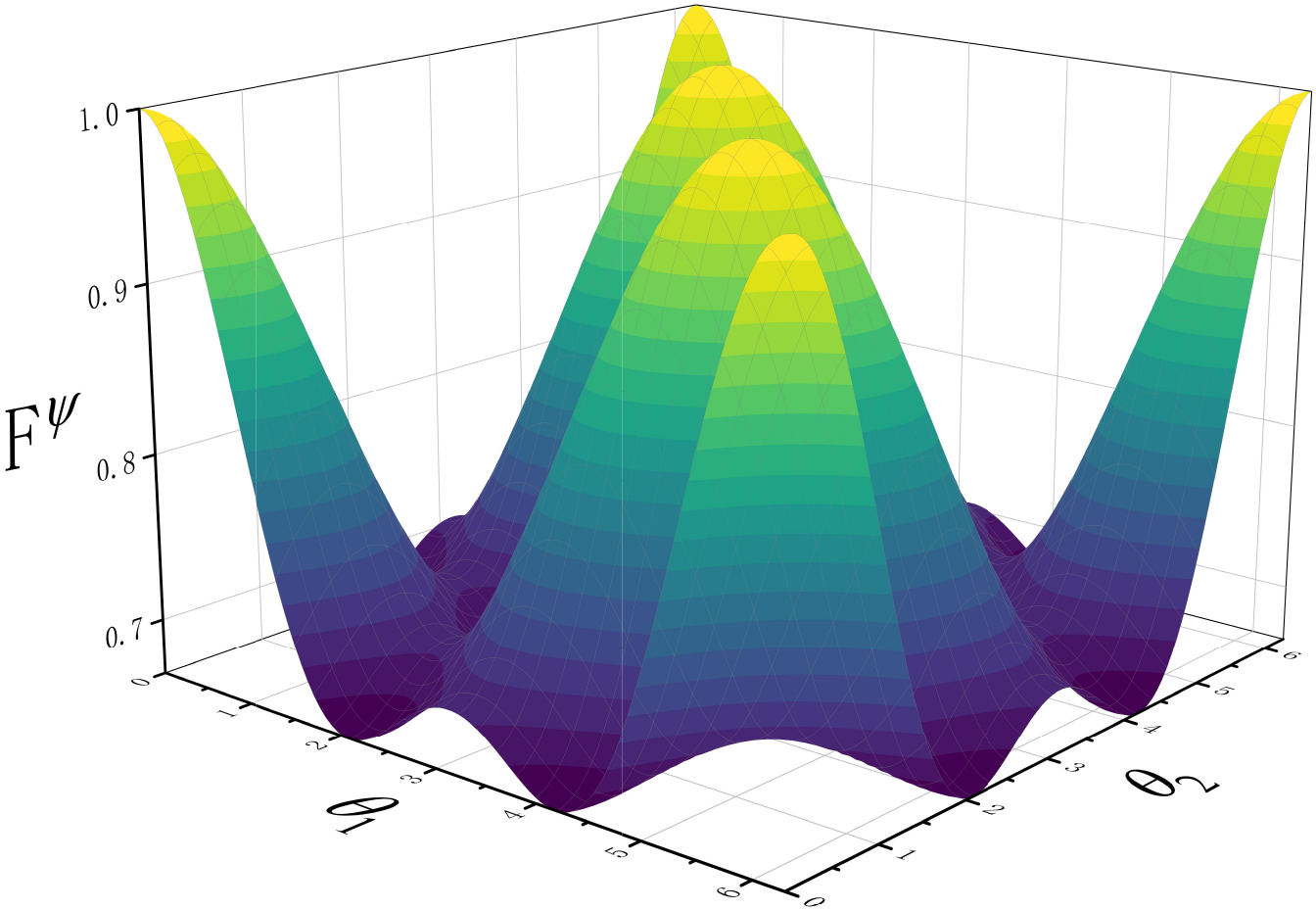}
\caption{The fidelity $F^{\psi}$ versus phase parameters $\theta_1,\theta_2$. It can be seen that for all $\theta_1,\theta_2$, $F^{\psi}\geq 2/3$ is satisfied. Additionally, for points $(\theta_1,\theta_2)=(0,\frac{2\pi}{3}),(0,\frac{4\pi}{3}),(\frac{2\pi}{3},0),(\frac{2\pi}{3},\frac{2\pi}{3}),(\frac{4\pi}{3},0),(\frac{4\pi}{3},\frac{4\pi}{3})$, $F^{\psi}= 2/3$.}
\label{Fig: F versus theta}
\end{figure*}

Here, we set the input state as $|\omega^{\mathrm{V}}\rangle=|\psi'^{\mathrm{V}}\rangle$, hence the coefficients $\alpha=(1+e^{\mathrm{i}\theta_1}+e^{\mathrm{i}\theta_2})/3,\beta=(1+e^{\mathrm{i}(\frac{2\pi}{3}+
\theta_1)}+e^{\mathrm{i}(\frac{4\pi}{3}+\theta_2)})/3,\gamma=(1+e^{\mathrm{i}(\frac{4\pi}{3}+
\theta_1)}+e^{\mathrm{i}(\frac{2\pi}{3}+\theta_2)})/3$. Suppose after the teleportation process, the device in Bob's side obtains state $\left|\Psi'^{\mathrm{B}}_{m n}\right\rangle$ which is the output state conditioned on Alice's BSM outcome $mn$. Then an additional operation $\mathcal{QFT}^{\dagger}_3$ is performed on state $\left|\Psi'^{\mathrm{B}}_{m n}\right\rangle$ such that the ultimate  recovered state is actually $\mathcal{QFT}^{\dagger}_3\left|\Psi'^{\mathrm{B}}_{m n}\right\rangle$. Noteworthy, the fidelity $F^\psi_{mn}$ between ultimate output state $\mathcal{QFT}^{\dagger}_3\left|\Psi'^{\mathrm{B}}_{m n}\right\rangle$ and initial input state $|\psi^{\mathrm{V}}\rangle$ is equivalent to the fidelity $F^{\psi'}_{mn}$ between state $\left|\Psi'^{B}_{m n}\right\rangle$ and state $|\psi'^{\mathrm{V}}\rangle$, i.e.,
\begin{equation}
F^\psi_{mn}=\langle \psi|(\mathcal{QFT}^{\dagger}_3|\Psi'_{mn}\rangle)(\langle \Psi'_{mn}|\mathcal{QFT}_3)|\psi\rangle=(\langle \psi|\mathcal{QFT}^{\dagger}_3)|\Psi'_{mn}\rangle\langle \Psi'_{mn}|(\mathcal{QFT}_3|\psi\rangle)=\langle \psi'|\Psi'_{mn}\rangle\langle \Psi'_{mn}|\psi'\rangle=F^{\psi'}_{mn}
\label{S-relation F F'}
\end{equation}
Thus $F^\psi_{mn}$ can be evaluate by the value of $F^{\psi'}_{mn}$. According to Table \ref{tab:table1}, for the state to teleport $|\psi'\rangle$, the teleportation fidelity of different measurement BSM outcomes $mn$ can be calculated as
\begin{equation}
\begin{aligned}
F^{\psi'}_{00}=F^{\psi'}_{10}=F^{\psi'}_{20}=&|\alpha|^2+|\beta|^2\\
&\frac{1}{9}[6+\cos(\theta_1)+\cos(\theta_2)+\cos(\theta_1-\theta_2)+\sqrt{3}\sin(\theta_1)-\sqrt{3}\sin(\theta_2)-\sqrt{3}\sin(\theta_1-\theta_2)],\\
F^{\psi'}_{01}=F^{\psi'}_{11}=F^{\psi'}_{21}=&|\alpha|^2+|\gamma|^2\\
=&\frac{1}{9}[6+\cos(\theta_1)+\cos(\theta_2)+\cos(\theta_1-\theta_2)-\sqrt{3}\sin(\theta_1)+\sqrt{3}\sin(\theta_2)+\sqrt{3}\sin(\theta_1-\theta_2)],\\
F^{\psi'}_{02}=F^{\psi'}_{12}=F^{\psi'}_{22}=&|\beta|^2+|\gamma|^2\\
=&\frac{2}{9}[3-\cos(\theta_1)-\cos(\theta_2)-\cos(\theta_1-\theta_2)].
\end{aligned}
\end{equation}
Furthermore, the probabilities of obtaining different BSM outcomes $mn$ are
\begin{equation}
\begin{aligned}
&p^{\psi'}_{00}=p^{\psi'}_{10}=p^{\psi'}_{20}=\frac{1}{54}[6+\cos(\theta_1)+\cos(\theta_2)+\cos(\theta_1-\theta_2)+\sqrt{3}\sin(\theta_1)-\sqrt{3}\sin(\theta_2)-\sqrt{3}\sin(\theta_1-\theta_2)],\\
&p^{\psi'}_{01}=p^{\psi'}_{11}=p^{\psi'}_{21}=\frac{1}{54}[6+\cos(\theta_1)+\cos(\theta_2)+\cos(\theta_1-\theta_2)-\sqrt{3}\sin(\theta_1)+\sqrt{3}\sin(\theta_2)+\sqrt{3}\sin(\theta_1-\theta_2),\\
&p^{\psi'}_{02}=p^{\psi'}_{12}=p^{\psi'}_{22}=\frac{1}{27}[3-\cos(\theta_1)-\cos(\theta_2)-\cos(\theta_1-\theta_2)].
\end{aligned}
\end{equation}
Hence, the teleportation fidelity of input state $|\psi'\rangle$ with two-dimensional Bell state as channel state is obtained as
\begin{equation}
F^{\psi'}=\sum_{mn}F^{\psi'}_{mn}p^{\psi'}_{mn}=\frac{1}{27}(21+2\cos(\theta_1-2\theta_2)+2\cos(2\theta_1-\theta_2)+2\cos(\theta_1+\theta_2)).
\end{equation}
Using Eq. \eqref{S-relation F F'}, one has the ultimate fidelity as
\begin{equation}
F^{\psi}=F^{\psi'}=\frac{1}{27}(21+2\cos(\theta_1-2\theta_2)+2\cos(2\theta_1-\theta_2)+2\cos(\theta_1+\theta_2)).
\end{equation}
As shown in Fig. \ref{Fig: F versus theta}, for all pair of parameters $\theta_1,\theta_2$ (except for $(\theta_1,\theta_2)=(0,\frac{2\pi}{3}),(0,\frac{4\pi}{3}),(\frac{2\pi}{3},0),(\frac{2\pi}{3},\frac{2\pi}{3}),(\frac{4\pi}{3},0),(\frac{4\pi}{3},\frac{4\pi}{3})$),  $F^{\psi}> 2/3$, which apparently leads to a judgement  that the above simulation process is a genuine three-dimensional QT based on the criterion $F_{\mathrm{tel}}^{\psi}>\frac{2}{3}$ proposed in Ref. \cite{luo2019quantum}. However, the channel state used in this simulation process is two-dimensional Bell state. Noteworthy, when $|\psi^{\mathrm{V}}\rangle=|\psi^{\mathrm{V}}_0\rangle=(|0^{\mathrm{V}}\rangle+|1^{\mathrm{V}}\rangle+|2^{\mathrm{V}}\rangle)/\sqrt{3}$, one has that $(\theta_1,\theta_2)=(0,0)$ and the ultimate fidelity $F^{\psi_0}=1$, the same as in the ideal three-dimensional QT scheme. These results indicates the fact that such a criterion cannot identify the simulation with lower-dimensional entanglement.

We also show that similar results also hold for the robustness criterion $\mu>0$ \cite{hu2020experimental}. For a simple instance, from Table \ref{tab:table1}, when Alice's BSM outcome is $mn=00$, the normalized output state of the teleportation process is
\begin{equation}
\left|\Psi'^{\mathrm{B}}_{00}\right\rangle=\frac{1}{\mathcal{N}}[(\frac{1+e^{\mathrm{i}\theta_1}+e^{\mathrm{i}\theta_2}}{3})|0^\mathrm{B}\rangle+(\frac{1+e^{\mathrm{i}(\frac{2\pi}{3}+
\theta_1)}+e^{\mathrm{i}(\frac{4\pi}{3}+\theta_2)}}{3})|1^\mathrm{B}\rangle]
\end{equation}
where $\mathcal{N}=|\alpha|^2+|\beta|^2$ is the normalization factor. By applying $\mathcal{QFT}^{\dagger}_3$, one has that
\begin{equation}
\begin{aligned}
|\Psi^{\mathrm{B}}_{00}\rangle=\mathcal{QFT}^{\dagger}_3 \left|\Psi'^{\mathrm{B}}_{00}\right\rangle=&\frac{1}{\sqrt{3}}\left(\begin{array}{ccc}
1 & 1 & 1 \\
1 & e^{\frac{4}{3} \pi \mathrm{i}} & e^{\frac{2}{3} \pi \mathrm{i}} \\
1 & e^{\frac{2}{3} \pi \mathrm{i}} & e^{\frac{4}{3} \pi \mathrm{i}}
\end{array}\right) \frac{1}{3\mathcal{N}}\left(\begin{array}{l}
1+e^{\mathrm{i}\theta_1}+e^{\mathrm{i}\theta_2} \\
1+e^{\mathrm{i}(\frac{2\pi}{3}+\theta_1)}+e^{\mathrm{i}(\frac{4\pi}{3}+\theta_2)} \\
0
\end{array}\right)\\
=&\frac{1}{3\sqrt{3}\mathcal{N}}
\left(\begin{array}{l}
2-e^{\mathrm{i}(\frac{4 \pi}{3}+\theta_1)}-e^{\mathrm{i}(\frac{2 \pi}{3}+\theta_2)} \\
-e^{\mathrm{i}\frac{2\pi}{3}}+2e^{\mathrm{i}\theta_1}-e^{\mathrm{i}(\frac{4\pi}{3}+\theta_2)} \\
-e^{\mathrm{i}\frac{4\pi}{3}}-e^{\mathrm{i}(\frac{2\pi}{3}+\theta_1)}+2e^{\mathrm{i}\theta_2}
\end{array}\right).
\end{aligned}
\end{equation}
For all cases except for $(\theta_1,\theta_2)=(\frac{2\pi}{3},\frac{4\pi }{3})$ (when $(\theta_1,\theta_2)=(\frac{2\pi}{3},\frac{4\pi }{3})$, $|\alpha|^2+|\beta|^2=0$, thus state $|\Psi^{\mathrm{B}}_{00}\rangle$ is meaningless because the probability of obtaining BSM outcome $00$ is zero), the ultimate output state $|\Psi^{\mathrm{B}}_{mn}\rangle$ is a genuine pure three-dimensional state that cannot be 
expressed as the mixture of the density matrix of qubits, 
which can lead to $\mu^{\psi}>0$ such that passing the robustness criterion \cite{hu2020experimental}. 

Hence, we conclude that both the criteria can be passed by using simulation strategy with only two-dimensional entanglement. These results reveal the fact the methods for demonstrating the teleportation of genuine three-dimensional states cannot be used to certify the resource. Instead, like the verification of nonclassical QT, the resource of genuine three-dimensional QT must also be verified through analyzing the teleportation data of input states that are tomographically complete, e.g. the 12 distinct states from the three-dimensional MUBs. 

\section{\label{AP:B}The maximal average fidelity of teleporting $d$-dimensional quantum states with using $(d'-1)$-dimensional entangled channel state}

In this section, we present the details about deriving the maximal average fidelity of teleporting $d$-dimensional quantum states via entangled channel state $\rho^{\mathrm{AB}}$ whose entanglement dimension is at most $d'-1$, i.e. $\bar{F}^{d^\prime-1}_{q,d}(\rho^{\mathrm{AB}})$.
Since $\rho^{\mathrm{AB}}$ has entanglement dimension at most $d^\prime-1$, there exists a decomposition $\rho^{\mathrm{AB}}=\sum_l p_l|\psi^\mathrm{AB}_l\rangle\langle\psi^\mathrm{AB}_l|$ where all $|\psi^\mathrm{AB}_l\rangle$ are pure entangled states of Schmidt rank at most $d^\prime-1$ \cite{terhal2000schmidt}. Under this decomposition, the normalized conditioned output state $\rho_{a \mid \omega_x}^{\mathrm{B}}$ can be written as
\begin{equation}
\rho_{a \mid \omega_x}^{\mathrm{B}}=\sum_l p_l \rho_{a \mid \omega_x}^{\mathrm{B},l},
\label{S-rhoB_mix}
\end{equation}
where $\rho_{a \mid \omega_x}^{\mathrm{B},l}=\operatorname{Tr}_{\mathrm{VA}}[(M_a^{\mathrm{VA}} \otimes \mathbb{I}^{\mathrm{B}}) \cdot(|\omega^{\mathrm{V}}_x\rangle\langle\omega^{\mathrm{V}}_x| \otimes |\psi^\mathrm{AB}_l\rangle\langle\psi^\mathrm{AB}_l|)]/p(a \mid \omega_x)$ is the output state corresponding to channel $|\psi^\mathrm{AB}_l\rangle\langle\psi^\mathrm{AB}_l|$. Using Eq. \eqref{S-rhoB_mix} and the concavity of fidelity function, we have
\begin{equation}
\bar{F}_{q, d}^{d^{\prime}-1}\left(\rho^{\mathrm{AB}}\right)=\sum_l p_l \bar{F}_{q, d}^{d^{\prime}-1}\left(\left|\psi_l^{\mathrm{AB}}\right\rangle\left\langle\psi_l^{\mathrm{AB}}\right|\right)\leq \bar{F}_{q, d}^{d^{\prime}-1}\left(\left|\psi_{l^{\prime}}^{\mathrm{AB}}\right\rangle\left\langle\psi_{l^{\prime}}^{\mathrm{AB}}\right|\right),
\label{S-decomp of fidelity}
\end{equation}
where $\left|\psi_{l^{\prime}}^{\mathrm{AB}}\right\rangle$ represents the pure channel state whose Schmidt rank is $d^\prime-1$. According to Ref. \cite{gong2024optimal}, the optimal average fidelity of teleporting $d$-dimensional states with given channel state has a form of 
\begin{equation}
\bar{F}_{q, d}^{d^{\prime}-1}\left(\left|\psi_{l^{\prime}}^{\mathrm{AB}}\right\rangle\left\langle\psi_{l^{\prime}}^{\mathrm{AB}}\right|\right)=\max _O\left\{\frac{1}{d}+\frac{d}{4(d+1)} \operatorname{Tr}\left(T^d_{\phi} T_{\rho} O\right)\right\}
\label{S-fidelity and correlation matrix}
\end{equation}
where $T_{\rho}$, which refers to the correlation matrix of $\left|\psi_{l^{\prime}}^{\mathrm{AB}}\right\rangle$, is a $(d^2-1)$-dimensional square matrix whose elements are defined by $t^{\rho}_{ij}=\operatorname{Tr}(\left|\psi_{l^{\prime}}^{\mathrm{AB}}\right\rangle\left\langle\psi_{l^{\prime}}^{\mathrm{AB}}\right|\lambda_i \otimes \lambda_j)$, $\{\lambda_i\}^{d^2-1}_{i=1}$ is the set of the $d$-dimensional generalized Gell-Mann matrices \cite{bertlmann2008bloch} (Also see Appendix \ref{Ap:C}). $T^d_{\phi}$ refers to the correlation matrix of Bell state $\left|\phi^d\right\rangle=\sum_{i=0}^{d-1}(1/\sqrt{d})|i\rangle|i\rangle$, and it's diagonal with elements
\begin{equation}
t^{\phi}_{ii}= \begin{cases}\frac{2}{d} & \text { for } i=k_1^2+2\left(k_2-1\right) \\ -\frac{2}{d} & \text { for } i=k_1^2+2\left(k_2-1\right)+1 \\ \frac{2}{d} & \text { for } i=\left(k_3+1\right)^2-1\end{cases}
\label{correlation matrix element of T00}
\end{equation}
where $k_1,k_2,k_3$ are integers satisfying $1 \leq k_1 \leq d-1,1 \leq k_2 \leq k_1, 1 \leq k_3 \leq d-1$. For a simple example when $d=3$,  $T^d_{\phi}=\operatorname{diag}\{2/3,-2/3,2/3,2/3,-2/3,2/3,-2/3,2/3\}$. Eq. \eqref{S-fidelity and correlation matrix} is maximized over all possible rotation $O$, which refers to the homomorphic matrix of a global local unitary operation $U$ performed by Bob (by global we mean that $U$ is fixed under all possible outcome $a$), For a given $U$, the elements of $O$ can be determined via $o_{ij}= \operatorname{Tr}\left(U \lambda_i U^{\dagger} \lambda_j\right)/2, \quad  \forall i,j=1,2,\ldots, d^2-1$ \cite{li2008upper}. 
Since channel $\left|\psi_{l^{\prime}}^{\mathrm{AB}}\right\rangle$ has Schmidt rank $d^\prime-1$, the general form of $\left|\psi_{l^{\prime}}^{\mathrm{AB}}\right\rangle$ is
\begin{equation}
|\psi_{l^{\prime}}^{\mathrm{AB}}\rangle=\left(U^{\mathrm{A}} \otimes U^{\mathrm{B}}\right) \sum_{i=0}^{d^{\prime}-2} \sqrt{\mu_i}|i^{\mathrm{A}}\rangle|i^{\mathrm{B}}\rangle,
\label{form of channel}
\end{equation}
where $\mu_i>0, \sum_{i=0}^{d^{\prime}-2} \mu_i =1$. Since local unitary transformations $U^{\mathrm{A}}\otimes U^{\mathrm{B}}$ won't contribute to the optimal average teleportation fidelity \cite{horodecki1999general}, we could simply set $U^\mathrm{A}=U^\mathrm{B}=I$. By derivation, correlation matrix $T_{\rho}$ has a highly regular structure as
\begin{equation}
t^{\rho}_{ij}= \begin{cases}2\sqrt{\mu_{k_1}\mu_{k_2-1}} & \text { for } i=j=k_1^2+2\left(k_2-1\right) \\ -2\sqrt{\mu_{k_1}\mu_{k_2-1}} & \text { for } i=j=k_1^2+2\left(k_2-1\right)+1 \\ 2\left(\sum^{\tilde{k}}_{s=0}\mu_s+g(k_3)g(k^\prime_3) \mu_{\tilde{k}+1}\right)/\sqrt{k_3 k^\prime_3 (k_3+1) (k^\prime_3+1)} & \text { for } i=\left(k_3+1\right)^2-1,j=\left(k^\prime_3+1\right)^2-1,i = j\\ 2\left(\sum^{\tilde{k}}_{s=0}\mu_s-g(\min(k_3,k^\prime_3))\mu_{\tilde{k}+1}\right)/\sqrt{k_3 k^\prime_3 (k_3+1) (k^\prime_3+1)} & \text { for } i=\left(k_3+1\right)^2-1,j=\left(k^\prime_3+1\right)^2-1,i\neq j\\
0 & \text { else } 
\end{cases},
\label{correlation matrix element of Trho}
\end{equation}
here in this formula, $k_1,k_2,k_3,k'_3$ are integers satisfying $1 \leq k_1 \leq d^\prime-2,1 \leq k_2 \leq k_1, 1 \leq k_3\leq d-1, 1 \leq k'_3\leq d-1$ and $\tilde{k}=\min\{k_3-1,k^\prime_3-1,d^\prime-3\}$. Function $g$ is defined by
\begin{equation}
g(x)= \begin{cases}1 & \text { for } x>d^\prime-2\\ x & \text { for } x \leq d^\prime-2  
\end{cases}.
\label{function g}
\end{equation}
The optimal value of Eq. \eqref{S-fidelity and correlation matrix} is obtained when $\operatorname{Tr}(T_\phi^d T_\rho O)$ is maximized. By calculation, one can see that matrix $\tilde{T}=T_\phi^d T_\rho$ is semidefinite (all eigenvalues of $\tilde{T}$ are non-negative), hence, $\max _O\{\operatorname{Tr}(\tilde{T} O)\}=\operatorname{Tr}(\tilde{T})$ (The optimal rotation $O$ is identity $I_{d^2-1}$). By tedious by straightforward derivation, it is obtained that
\begin{small}
\begin{equation}
\begin{aligned}
\operatorname{Tr}(\tilde{T})&=\frac{2}{d}\left[4\sum^{d^\prime-2}_{k_1=1}\sum^{k_1}_{k_2=1}\sqrt{\mu_{k_1}\mu_{k_2-1}}+2(1-\frac{1}{d})\sum^{d^\prime-2}_{s=0}\mu_s\right]\\
&\leq \frac{2}{d}\left[4\sum^{d^\prime-2}_{k_1=1}\sum^{k_1}_{k_2=1}\frac{\mu_{k_1}+\mu_{k_2-1}}{2}+2(1-\frac{1}{d})\sum^{d^\prime-2}_{s=0}\mu_s\right].
\label{TrT}
\end{aligned}
\end{equation}
\end{small}
The maximum of Eq. \eqref{TrT} is achieved when $\mu_0=\mu_1=...=\mu_{d^\prime-2}=1/d^\prime-1$. Subsequently, one arrives at
\begin{equation}
    \max \operatorname{Tr}(\tilde{T})=\frac{4[d(d^\prime-1)-1]}{d^2}.
    \label{TrT2}
\end{equation}
Substituting Eq. \eqref{TrT2} into Eq. \eqref{S-fidelity and correlation matrix}, the maximal average fidelity of teleporting $d$-dimensional states with using $(d^\prime-1)$-dimensional entangled state is 
\begin{equation}
    \bar{F}_{q, d}^{d^{\prime}-1}=\frac{d^\prime}{d+1}.
    \label{S-result1}
\end{equation}

Eq.\eqref{S-result1} can also be derived from a rather simpler way. For a given bipartite channel state $\rho^{\mathrm{AB}}$ defined in Hilbert space $\mathcal{H}^d \otimes \mathcal{H}^d$, the maximal average fidelity of teleporting $d$-dimensional states is connected with the fully entangled fraction (FEF) $f(\rho^{\mathrm{AB}})$ of the channel state, i.e. $\bar{F}_{max}=[f(\rho^{\mathrm{AB}})d+1]/(d+1)$ \cite{horodecki1999general}. And it is proved that the FEF of those states with Schmidt number at most $d'-1$ is upper bounded by $f(\rho^{\mathrm{AB}})\leq (d'-1)/d$ \cite{terhal2000schmidt}, hence leading to Eq. \eqref{S-result1}. Although such derivation is concise and direct with the help of the notion of FEF, it is essential to proceed from the quantum teleportation process itself (see Eq. \eqref{S-fidelity and correlation matrix}). Moreover, the consistency of conclusions derived from different methods confirms the correctness of Eq. \eqref{S-result1}.

\section{\label{Ap:C}The generalized Gell-Mann matrix}

The generalized Gell-Mann matrices (GGM) are higher-dimensional extensions of the Pauli matrices (for qubits) and the Gell-Mann matrices (for qutrits), they are the standard generators for group $\mathrm{SU}(d)$. The GGM are defined as three different types of matrices: 

i) $d(d-1)/2$ symmetric GGM. For integer $k_1$ and $k_2$ that satisfies $1\leq k_1 \leq d-1, 1\leq k_2 \leq k_1$, symmetric GGM can be expressed as
\begin{equation}
\lambda_{k^2_1+2(k_2-1)}=|k_2-1\rangle\langle k_1|+|k_1\rangle\langle k_2-1|
\end{equation}

ii) $d(d-1)/2$ asymmetric GGM. For integer $k_1$ and $k_2$ that satisfies $1\leq k_1 \leq d-1, 1\leq k_2 \leq k_1$, asymmetric GGM can be expressed as
\begin{equation}
\lambda_{k^2_1+2(k_2-1)+1}=-\operatorname{i}|k_2-1\rangle\langle k_1|+\operatorname{i}|k_1\rangle\langle k_2-1|
\end{equation}

iii) $d-1$ diagonal GGM. For integer $k_3$ that satisfies $1\leq k_3 \leq d-1$, diagonal GGM can be expressed as
\begin{equation}
\lambda_{(k_3+1)^2-1}=\sqrt{\frac{2}{k_3(k_3+1)}}(\sum_{l=0}^{k_3-1}|l\rangle\langle l|-k_3|k_3\rangle\langle k_3|)
\end{equation}

For the case of $d=2$, we obtain the standard pauli matrices for $\text{SU}(2)$ as
\begin{equation}
\lambda_1=\left(\begin{array}{cc}
0 & 1 \\
1 & 0
\end{array}\right) ; \lambda_2=\left(\begin{array}{cc}
0 & -\operatorname{i} \\
\operatorname{i} & 0
\end{array}\right) ; \lambda_3=\left(\begin{array}{cc}
1 & 0 \\
0 & -1
\end{array}\right)
\end{equation}

For the case of $d=3$, we obtain the Gell-Mann matrices for $\text{SU}(3)$ as
\begin{small}
\begin{equation}
\begin{array}{llll}
\lambda_1=\left(\begin{array}{lll}
0 & 1 & 0 \\
1 & 0 & 0 \\
0 & 0 & 0
\end{array}\right); & \lambda_2=\left(\begin{array}{ccc}
0 & -\operatorname{i}  & 0 \\
\operatorname{i}  & 0 & 0 \\
0 & 0 & 0
\end{array}\right); & \lambda_3=\left(\begin{array}{ccc}
1 & 0 & 0 \\
0 & -1 & 0 \\
0 & 0 & 0
\end{array}\right); & \lambda_4=\left(\begin{array}{lll}
0 & 0 & 1 \\
0 & 0 & 0 \\
1 & 0 & 0
\end{array}\right);\\
\lambda_5=\left(\begin{array}{ccc}
0 & 0 & -\operatorname{i}  \\
0 & 0 & 0 \\
\operatorname{i}  & 0 & 0
\end{array}\right); & \lambda_6=\left(\begin{array}{lll}
0 & 0 & 0 \\
0 & 0 & 1 \\
0 & 1 & 0
\end{array}\right);&\lambda_7=\left(\begin{array}{ccc}
0 & 0 & 0 \\
0 & 0 & -\operatorname{i}  \\
0 & \operatorname{i}  & 0
\end{array}\right); & \lambda_8=\frac{1}{\sqrt{3}}\left(\begin{array}{ccc}
1 & 0 & 0 \\
0 & 1 & 0 \\
0 & 0 & -2
\end{array}\right) .
\end{array}
\end{equation}    
\end{small}

\section{\label{Ap:D}Two-dimensional maximally entangled state is sufficient to achieve an average fidelity of 3/4 in standard three-dimensional QT scheme}

In this section, we show that two-dimensional Bell state is sufficient to obtain an average fidelity of $\frac{3}{4}$ in standard three-dimensional QT scheme.

The input state is $|\omega^\mathrm{V}\rangle=\alpha|0\rangle+\beta|1\rangle+\gamma|2\rangle$ satisfying the normalization condition $\left|\alpha\right|^2+\left|\beta\right|^2+\left|\gamma\right|^2=1$. The channel state used is two-dimensional Bell state $|\varphi^{\mathrm{AB}}\rangle=\frac{1}{\sqrt{2}}(|0^{\mathrm{A}}\rangle|0^{\mathrm{B}}\rangle+|1^{\mathrm{A}}\rangle|1^{\mathrm{B}}\rangle)$. The detailed scheme can be found in Appendix \ref{Ap:A}

According to Table \ref{tab:table1}, the teleportation fidelity of given input state $|\omega^{\mathrm{V}}\rangle$ can be obtained by the weighted average over all outcomes of BSM,
\begin{equation}
F^\omega=\sum_{mn}F_{mn}p_{mn}=|\alpha|^4+|\beta|^4+|\gamma|^4+|\alpha|^2|\beta|^2+|\alpha|^2|\gamma|^2+|\beta|^2|\gamma|^2
\end{equation}

By setting $x=|\alpha|^2,y=|\beta|^2,z=|\gamma|^2=1-x-y$, the average teleportation fidelity over all pure three-dimensional states can be simplified as
\begin{equation}
\bar{F}=\int_\omega F^{\omega}\mathrm{d}M(\omega)=\frac{\int\int_D(x^2+y^2+xy-x-y+1) dxdy}{\int\int_Ddxdy},
\label{S-Fbar}
\end{equation}
where $D=\{(x, y) \mid 0 \leq x \leq 1,0 \leq y \leq 1-x\}$.
By straightforward calculation, Eq. \eqref{S-Fbar} becomes
\begin{equation}
\bar{F}=\frac{\int^1_0\int^{1-x}_0(x^2+y^2+xy-x-y+1) dydx}{\int^1_0\int^{1-x}_0dydx}=\frac{3}{8}/\frac{1}{2}=\frac{3}{4},
\label{S-Fbar2}
\end{equation}
We conclude that, by using two-dimensional maximally entangled state as resource, the average teleportation fidelity can be sufficiently realized as $\frac{3}{4}$ within standard three-dimensional quantum teleportation scheme.

To fully characterize the teleportation performance in experiment, one can choose the minimal tomographically-complete set of input states, which is composed of all 12 qutrit states $|\omega^{\mathrm{V}}_i\rangle$($i=1,2,3,...,12$) from 4 mutually unbiased bases for three-dimensional system \cite{luo2019quantum}. That is, the average fidelity of these 12 qutrit states is the same as that averaging over the whole pure-state space \cite{Ivonovic1981GeometricalDO}. The 12 qutrit states are
\begin{equation}
\begin{array}{lll}
|\omega^{\mathrm{V}}_1\rangle=|0\rangle,&|\omega^{\mathrm{V}}_2\rangle=|1\rangle,&|\omega^{\mathrm{V}}_3\rangle=|2\rangle,\\
|\omega^{\mathrm{V}}_4\rangle=\frac{1}{\sqrt{3}}(|0\rangle+|1\rangle+|2\rangle),&|\omega^{\mathrm{V}}_5\rangle=\frac{1}{\sqrt{3}}(|0\rangle+e^{\frac{2 \pi \mathrm{i}}{3}}|1\rangle+e^{\frac{4 \pi \mathrm{i}}{3}}|2\rangle),&|\omega^{\mathrm{V}}_6\rangle=\frac{1}{\sqrt{3}}(|0\rangle+e^{\frac{4 \pi \mathrm{i}}{3}}|1\rangle+e^{\frac{2 \pi \mathrm{i}}{3}}|2\rangle)\\
|\omega^{\mathrm{V}}_7\rangle=\frac{1}{\sqrt{3}}(|0\rangle+e^{\frac{2 \pi \mathrm{i}}{3}}|1\rangle+e^{\frac{2 \pi \mathrm{i}}{3}}|2\rangle),&|\omega^{\mathrm{V}}_8\rangle=\frac{1}{\sqrt{3}}(e^{\frac{2 \pi \mathrm{i}}{3}}|0\rangle+|1\rangle+e^{\frac{2 \pi \mathrm{i}}{3}}|2\rangle),&|\omega^{\mathrm{V}}_9\rangle=\frac{1}{\sqrt{3}}(e^{\frac{2 \pi \mathrm{i}}{3}}|0\rangle+e^{\frac{2 \pi \mathrm{i}}{3}}|1\rangle+|2\rangle)\\
|\omega^{\mathrm{V}}_{10}\rangle=\frac{1}{\sqrt{3}}(|0\rangle+e^{\frac{4 \pi \mathrm{i}}{3}}|1\rangle+e^{\frac{4 \pi \mathrm{i}}{3}}|2\rangle),&|\omega^{\mathrm{V}}_{11}\rangle=\frac{1}{\sqrt{3}}(e^{\frac{4 \pi \mathrm{i}}{3}}|0\rangle+|1\rangle+e^{\frac{4 \pi \mathrm{i}}{3}}|2\rangle),&|\omega^{\mathrm{V}}_{12}\rangle=\frac{1}{\sqrt{3}}(e^{\frac{4 \pi \mathrm{i}}{3}}|0\rangle+e^{\frac{4 \pi \mathrm{i}}{3}}|1\rangle+|2\rangle).
\end{array}
\end{equation}

When the input state is $|\omega^{\mathrm{V}}_1\rangle=|0\rangle$, one has $|\alpha|^2=1,|\beta|^2=0,|\gamma|^2=0$. From Table \ref{tab:table1}, the teleportation fidelitys conditioned on Alice's outcomes $mn$ are $f^{1}_{00}=f^{1}_{10}=f^{1}_{20}=f^{1}_{01}=f^{1}_{11}=f^{1}_{21}=1$ with probability $p^{1}_{00}=p^{1}_{10}=p^{1}_{20}=p^{1}_{01}=p^{1}_{11}=p^{1}_{21}=\frac{1}{6}$ (the subscript 1 referring to $|\omega^{\mathrm{V}}_1\rangle$). And since $p_{02}=p_{12}=p_{22}=0$, the projection events that systems V and A are projected onto  $\left|\phi_{02}^{\mathrm{VA}}\right\rangle, \left|\phi_{12}^{\mathrm{VA}}\right\rangle$ and $\left|\phi_{22}^{\mathrm{VA}}\right\rangle$ will not happen. Hence, the teleportation fidelity of teleporting state $|\omega^{\mathrm{V}}_1\rangle$ is $f^1=\sum_{mn}p^1_{mn}f^1_{mn}=1$. When the input state is $|\omega^{\mathrm{V}}_2\rangle=|1\rangle$, one has $|\alpha|^2=0,|\beta|^2=1,|\gamma|^2=0$. Then it also be obtained that $f^{2}_{00}=f^{2}_{10}=f^{2}_{20}=f^{2}_{02}=f^{2}_{12}=f^{2}_{22}=1$ with probability $p^{2}_{00}=p^{2}_{10}=p^{2}_{20}=p^{2}_{02}=p^{2}_{12}=p^{2}_{22}=\frac{1}{6}$, Hence, the teleportation fidelity of teleporting state $|\omega^{\mathrm{V}}_2\rangle$ is $f^2=\sum_{mn}p^2_{mn}f^2_{mn}=1$. Similar result can also be obtained for input $|\omega^{\mathrm{V}}_3\rangle=|2\rangle$, i.e., $f^3=1$.

For the rest 9 input states $|\omega^{\mathrm{V}}_4\rangle\sim |\omega^{\mathrm{V}}_{12}\rangle$, the parameters $|\alpha|^2,|\beta|^2,|\gamma|^2$ are the same as $|\alpha|^2=|\beta|^2=|\gamma|^2=\frac{1}{3}$. Hence, for $i=4,5,6,...,12$, the following is true, $f^{i}_{00}=f^{i}_{10}=f^{i}_{01}=f^{i}_{11}=f^{i}_{21}=f^{i}_{20}=f^{i}_{02}=f^{i}_{12}=f^{i}_{22}=\frac{2}{3}$ and $p^{i}_{00}=p^{i}_{10}=p^{i}_{20}=p^{i}_{01}=p^{i}_{11}=p^{i}_{21}=p^{i}_{02}=p^{i}_{12}=p^{i}_{22}=\frac{1}{9}$. Consequently the corresponding teleportation fidelity can be calculated as $f^i=\sum_{mn}p^i_{mn}f^i_{mn}=\frac{2}{3}$.

Since the input states are distributed equally, the average teleportation fidelity is
\begin{equation}
\bar{f}=\frac{1}{12}\sum_i f^{i}=\frac{1}{12}(1\times3+\frac{2}{3}\times9)=\frac{3}{4},
\end{equation}
exactly the same as Eq. \eqref{S-Fbar2}.

\section{\label{Ap:E}The fidelity-based criterion is feasible when Alice performs partial BSM}

In Appendix \ref{AP:B}, we derived that the maximal average teleportation fidelity using $(d^\prime-1)$-dimensional entanglement resource is $d^\prime/(d+1)$, i.e. Eq. \eqref{S-result1}. This result also shows that optimal $(d^\prime-1)$-dimensional entangled state is $|\psi_{d^{\prime}-1}^{\mathrm{AB}}\rangle=\sum_{i=0}^{d^{\prime}-2} (1/\sqrt{d^\prime-1})|i^{\mathrm{A}}\rangle|i^{\mathrm{B}}\rangle$. This conclusion is obtained under the assumption that Alice executes a full BSM. In this section, we show that Eq. \eqref{S-result1} is also correct in the scenario where only partial BSM is available. 

Suppose the input state is $|\omega^\mathrm{V}\rangle=\sum_{k=0}^{d-1} \alpha_k|k\rangle$ satisfying the normalization condition $\sum_{k=0}^{d-1}\left|\alpha_k\right|^2=1$. The channel state used is $|\psi_{d^{\prime}-1}^{\mathrm{AB}}\rangle$. The measurement performed by Alice is partial BSM $\{|\phi_d^{\mathrm{VA}}\rangle\langle\phi_d^{\mathrm{VA}}|, \mathbb{I}-|\phi_d^{\mathrm{VA}}\rangle\langle\phi_d^{\mathrm{VA}}|\}$, where $|\phi_d^{\mathrm{VA}}\rangle=\sum_{i=0}^{d-1} (1/\sqrt{d})|i^{\mathrm{V}}\rangle|i^{\mathrm{A}}\rangle$. The main goal is to derive the maximal average teleportation fidelity conditioned on the case that Alice's projection is $|\phi_d^{\mathrm{VA}}\rangle\langle\phi_d^{\mathrm{VA}}|$. The combined state of system VAB is 
\begin{equation}
|\Psi^{\mathrm{VAB}}\rangle=|\omega^{\mathrm{V}}\rangle \otimes|\psi_{d^\prime-1}^{\mathrm{AB}}\rangle=\frac{1}{\sqrt{d^\prime-1}}[\sum^{d-1}_{k=0}\alpha_k|k^\mathrm{V}\rangle(\sum^{d^\prime-2}_{i=0}|i^\mathrm{A}\rangle|i^\mathrm{B}\rangle)].
\end{equation}
The projection of system VA onto $|\phi_d^{\mathrm{VA}}\rangle$ collapses system B into unnormalized state
\begin{equation}
|\tilde{\psi}^{\mathrm{B}}_{\text{out}}\rangle=\sqrt{p}|\psi^{\mathrm{B}}_{\text{out}}\rangle=\frac{\sqrt{\sum^{d'-2}_{l=0}|\alpha_k|^2}}{\sqrt{d(d^\prime-1)}}\sum^{d^\prime-1}_{k=0}\frac{\alpha_k}{\sqrt{\sum^{d'-2}_{l=0}|\alpha_k|^2}}|k^{\mathrm{B}}\rangle.
\label{S3-2}
\end{equation}
The fidelity function of normalized output state $|\psi^{\mathrm{B}}\rangle$ is
\begin{equation}
F_\omega=\langle\omega|\psi_{out}\rangle\langle\psi_{out}|\omega\rangle=\sum^{d'-2}_{k=0}|\alpha_k|^2=1-\sum^{d-1}_{k=d'-1}|\alpha_k|^2=1-Y,
\label{S3-3}
\end{equation}
where $Y=\sum^{d-1}_{k=d'-1}|\alpha_k|^2$. According to Eq. \eqref{S3-2} and \eqref{S3-3}, one has that when the input state is $|\omega^{\mathrm{V}}\rangle$, the teleportation fidelity with the successful projection is obtained as $F_\omega$ whose probability is $p(F_\omega|Y)=(1-Y)/d(d'-1)$.
The goal is to calculate the average teleportation fidelity $\bar{F}$ conditioned on successful projection event $I=1$ ($I$ could be regard as the partial BSM, and $I=1$ refers to the event that the measurement successfully projects system VA onto $|\phi^{\mathrm{VA}}_d\rangle$), which can be written as the form of conditioned expectation value of variable $F_\omega$,
\begin{equation}
\bar{F}=\mathbb{E}[F_\omega|I=1]=\sum_Y(1-Y)p(F_\omega=1-Y|I=1),
\label{S3-4}
\end{equation}
where $P(F_\omega=1-Y|I=1)$ denotes the corresponding conditioned probability. Using Bayes' theorem, one obtain
\begin{equation}
\bar{F}=\sum_Y(1-Y)\frac{p(F_\omega=1-Y,I=1)}{p(I=1)}.
\label{S3-5}
\end{equation}
$p(F_\omega=1-Y,I=1)$ is the probability of $F_\omega=1-Y$ which combines two aspects, the probability of successful projection conditioned on input $|\omega^{\mathrm{V}}\rangle$ and the probability of choosing $|\omega^{\mathrm{V}}\rangle$ as input i.e. $p(F_\omega=1-Y,I=1)=p(F_\omega=1-Y|Y)p(Y)$. $P(I=1)$ is the probability of successful projection regardless of parameter $M$, i.e. $p(I=1)=\mathbb{E}(p(F_\omega=1-Y|Y))$. Hence, Eq. \eqref{S3-5} becomes
\begin{equation}
\bar{F}=\sum_Y(1-Y)\frac{p(F_\omega=1-Y|Y)p(Y)}{\mathbb{E}(p(F_\omega=1-Y|Y))}=\frac{\mathbb{E}[(1-Y)^2]}{\mathbb{E}(1-Y)}=\frac{1-2\mathbb{E}(Y)+\mathbb{E}(Y^2)}{1-\mathbb{E}(Y)}.
\label{S3-5-1}
\end{equation}
To calculate $\mathbb{E}(Y)$ and $\mathbb{E}(Y^2)$, one can use that fact that the square amplitudes $\{\left|\alpha_k\right|^2\}_{k=0}^{d-1}$ obey a Dirichlet distribution under the uniform Haar measure \cite{PhysRevResearch.3.043145}. Specifically,
\begin{equation}
\left(|\alpha_0|^2, |\alpha_1|^2, \ldots, |\alpha_{d-1}|^2\right) \sim \operatorname{Dirichlet}(\underbrace{1,1, \ldots, 1}_{d \text { terms }}) .
\end{equation}
The Dirichlet $(1,1, \ldots, 1)$ distribution corresponds to the uniform distribution over the $(d-1)$ simplex, ensuring all pure quantum states are equally likely in Hilbert space. For any component $|\alpha_{k}|^2$, the following is true,
\begin{equation}
\mathbb{E}(|\alpha_{k}|^2)=\frac{1}{d},\mathbb{E}\left(|\alpha_{k}|^4\right)=\frac{2}{d(d+1)}, \operatorname{Var}(|\alpha_{k}|^2)=\frac{d-1}{d^2(d+1)}, \operatorname{Cov}(\left|\alpha_i\right|^2,\left|\alpha_j\right|^2)=-\frac{1}{d^2(d+1)} \quad(i \neq j).
\label{S3-8}
\end{equation}
Using Eq. \eqref{S3-8}, one has that for variable $Y=\sum^{d-1}_{k=d'-1}|\alpha_k|^2$, 
\begin{equation}
\begin{aligned}
&\mathbb{E}(Y)=\sum^{d-1}_{k=d'-1} \cdot \mathbb{E}(\left|\alpha_k\right|^2)=\frac{d-(d'-1)}{d},\\
&\operatorname{Var}(Y)=\sum_{k=d^{\prime}-1}^{d-1} \operatorname{Var}(\left|\alpha_k\right|^2)+2 \sum_{i<j} \operatorname{Cov}(\left|\alpha_i\right|^2,\left|\alpha_j\right|^2)=\frac{(d-d'+1)(d'-1)}{d^2(d+1)},\\
&\mathbb{E}\left(Y^2\right)=\operatorname{Var}(Y)+[\mathbb{E}(Y)]^2=\frac{\left(d-d^{\prime}+1\right)\left(d+2-d^{\prime}\right)}{d(d+1)}.
\end{aligned}
\label{S3-9}
\end{equation}
Substituting Eq. \eqref{S3-9} into Eq. \eqref{S3-5-1}, we have
\begin{equation}
    \bar{F}=\frac{d'}{d+1}.
\end{equation}
Hence, the average fidelity of teleporting $d$-dimensional states using $(d^\prime-1)$-dimensional maximally entangled state $\left|\psi_{d^{\prime}-1}^{\mathrm{AB}}\right\rangle$ with partial BSM is $d'/(d+1)$. The derivation is also valid for cases where the system VA is projected onto the other $d^2-1$ Bell states. We conclude that the fidelity-based criterion is also feasible in HDQT scenario where Alice executes partial BSM, which is experiment-friendly.

\section{\label{Ap:F}All channel states with $d'$-dimensional entanglement can demonstrate genuine $d'$-dimensional quantum teleportation}

Here we prove that when (i) at least one of Alice's measurement operators corresponds to a projection onto a maximally entangled state and (ii) the inputs $|\omega^{\mathrm{V}}_x\rangle$ are tomographically complete, $R^{d^{\prime}}_{\text{tel}}(\sigma_{a \mid \omega_x}^{\mathrm{B}})$ is proportional to $R^{d^{\prime}}_{\text{ent}}(\rho^{\mathrm{AB}})$, the generalized genuine $d$-dimensional entanglement robustness of the state $\rho^{\mathrm{AB}}$. And if Alice's measurement is full BSM, $R^{d^{\prime}}_{\text{tel}}(\sigma_{a \mid \omega_x}^{\mathrm{B}})=R^{d^{\prime}}_{\text{ent}}(\rho^{\mathrm{AB}})$. 

To start, we write the SDP program (Eq. \eqref{SDP-GHTR}) in an equivalent form,
\begin{subequations}\label{S4-1}
    \begin{align}
        & \text { given }\{\sigma_{a \mid \omega_x}^{\mathrm{B}}\}_{a, x} \label{S4-1-1} \\
        & R^{d^{\prime}}_{\text{tel}}(\sigma_{a \mid \omega_x}^{\mathrm{B}})=\min _{ r_a, \{\tilde{N}_a^{\mathrm{VB}}\},\{M_a^{\mathrm{VB}}\}} r \label{S4-1-2} \\
        & \text { such that } \notag \\
        & \sigma_{a \mid \omega_x}^{\mathrm{B}}+\operatorname{Tr}_{\mathrm{V}}\left[r_a \tilde{N}_a^{\mathrm{VB}}\left(\omega_x^{\mathrm{V}} \otimes \mathbb{I}^{\mathrm{B}}\right)\right]=\operatorname{Tr}_{\mathrm{V}}\left[M_a^{\mathrm{VB}}\left(\omega_x^{\mathrm{V}} \otimes \mathbb{I}^{\mathrm{B}}\right)\right] \quad \forall x, a,\label{S4-1-3}\\
        & \sum_a r_a \tilde{N}_a^{\mathrm{VB}}=\mathbb{I}^{\mathrm{V}} \otimes r\tilde{\rho}^{\mathrm{B}}_N,\label{S4-1-4}\\
        & \sum_a M_a^{\mathrm{VB}}=\mathbb{I}^{\mathrm{V}} \otimes(\sum_a \sigma_{a \mid \omega_1}^{\mathrm{B}}+r\tilde{\rho}^{\mathrm{B}}_N),\label{S4-1-5}\\
        & M_a^{\mathrm{VB}}\in S_{d'-1} \quad \forall a, \label{S4-1-6}\\
        %& (\mathbb{I}^{\mathrm{V}} \otimes \Lambda^{\mathrm{B}}_{d^\prime-1}) M_a^{\mathrm{VB}} \geq 0 \quad \forall a, \label{S4-1-6}\\
        & \tilde{N}_a^{\mathrm{VB}} \in {S_1} \quad \forall a,\label{S4-1-7}\\
        & \text{Tr}(\tilde{\rho}_N^{\mathrm{B}}) = 1,
        \label{S4-1-8}
    \end{align}
\end{subequations}
where $\tilde{\rho}_N^{\mathrm{B}}$ refers to normalized teleportation data corresponding to noise, and $\omega^{\mathrm{V}}_x$ represents the density matrix of input state which is not necessary to be pure. $S_{k}$ refers to the set of operators with Schmidt number at most $k$, and especially, $S_1$ is the set of separable operators. Then, similarly, we also rewrite the generalized $d'$-dimensional entanglement robustness ($\mathrm{GHER}_{d'}$) of the channel state $\rho^{\mathrm{AB}}$,
\begin{subequations}\label{S4-2}
    \begin{align}
        & \text { given }\rho^{\mathrm{AB}} \label{S4-2-1} \\
        & R^{d^{\prime}}_{\text{ent}}(\rho^{\mathrm{AB}})=\min _{t,\tilde{N}^{\mathrm{AB}},M^{\mathrm{AB}}} t  \label{S4-2-2} \\
        & \text { such that } \notag \\
        & \rho^{\mathrm{AB}}+t \tilde{N}^{\mathrm{AB}}=M^{\mathrm{AB}},\label{S4-2-3}\\
        %& (\mathbb{I}^{\mathrm{A}} \otimes \Lambda^{\mathrm{B}}_{d^\prime-1}) M^{\mathrm{AB}} \geq 0,\label{S4-2-4}\\
        & M^{\mathrm{AB}}\in S_{d'-1} , \label{S4-2-4}\\
        & \tilde{N}^{\mathrm{AB}} \in {S_1}.\label{S4-2-5}\\
        & \text{Tr}\tilde{N}^{\mathrm{AB}} =1.\label{S4-2-6}
    \end{align}
\end{subequations}
Any non-negative value of $R^{d^{\prime}}_{\text{ent}}$ indicates that $\rho^{\mathrm{AB}}$ is at least $d'$-dimensional entangled.

The main idea of this proof is to show that the optimization problems \eqref{S4-1} and \eqref{S4-2} have the same structure with parameter $t$ is proportional to $r$ if condition (i) and (ii) are satisfied.
Starting with the case $a=1$, i.e. $M_1^{\mathrm{VA}}=|\phi_d^{\mathrm{VA}}\rangle\langle\phi_d^{\mathrm{VA}}|$, and suppose the channel state used is $\rho^{\mathrm{AB}}$, we can see that the left-hand-side of Eq. \eqref{S4-1-3} can be written as
\begin{equation}
\sigma_{1 \mid \omega_x}^{\mathrm{B}}+\operatorname{Tr}_{\mathrm{V}}\left[r_1 \tilde{N}_1^{\mathrm{VB}}\left(\omega_x^{\mathrm{V}} \otimes \mathbb{I}^{\mathrm{B}}\right)\right]=\text{Tr}_\mathrm{VA}\left\{(|\phi_d^{\mathrm{VA}}\rangle\langle\phi_d^{\mathrm{VA}}|\otimes \mathbb{I}^{\mathrm{B}})\cdot \left[\omega^{\mathrm{V}}_x \otimes (\rho^{\mathrm{AB}}+r_1\tilde{\rho}^{\mathrm{AB}}_{N,1})\right]\right\}
\label{S4-3}
\end{equation}
where normalized separable operator $\tilde{\rho}_{N,1}^{\mathrm{AB}}$ refers to the noisy channel state corresponding to $\tilde{N}_1^{\mathrm{VB}}$. Using the identity $\operatorname{Tr}_{\mathrm{A}}\left[\left(|\phi_d^{\mathrm{VA}}\rangle\langle\phi_d^{\mathrm{VA}}| \otimes \mathbb{I}^{\mathrm{B}}\right)\left(\mathbb{I}^{\mathrm{V}} \otimes \tau^{\mathrm{AB}}\right)\right]=(1/d)\left[\tau^{\mathrm{VB}}\right]^{T _\mathrm{V}}$ where $\tau^{\mathrm{VB}}=\tau^{\mathrm{AB}}$ \cite{cavalcanti2017all}, one has that $\tilde{N}_1^{\mathrm{VB}}=(\tilde{\rho}_{N,1}^{\mathrm{VB}})^{T_\mathrm{V}}/d=(\tilde{\rho}_{N,1}^{\mathrm{AB}})^{T_\mathrm{V}}/d$. Then Eq. \eqref{S4-3} becomes
\begin{equation}
\sigma_{1 \mid \omega_x}^{\mathrm{B}}+\operatorname{Tr}_{\mathrm{V}}\left[r_1 \tilde{N}_1^{\mathrm{VB}}\left(\omega_x^{\mathrm{V}} \otimes \mathbb{I}^{\mathrm{B}}\right)\right]=\frac{1}{d}\text{Tr}_{\mathrm{V}}\left[(\rho^{\mathrm{VB}}+r_1 \tilde{\rho}_{N,1}^{\mathrm{VB}})^{T_\mathrm{V}}\cdot (\omega^{\mathrm{V}}_x \otimes \mathbb{I}^{\mathrm{B}})\right]
\label{S4-4}\end{equation}
Combining Eq. \eqref{S4-4} and \eqref{S4-1-3}, one obtains
\begin{equation}
\frac{1}{d}(\rho^{\mathrm{VB}}+r_1 \tilde{\rho}_{N,1}^{\mathrm{VB}})=(M^{\mathrm{VB}}_1)^{T_\mathrm{V}}.
\label{S4-5}
\end{equation}
%This can equivalently be written as
%\begin{equation}
%\rho^{\mathrm{VB}}+r_1 %\tilde{\rho}_1^{\mathrm{VB}}=\Omega_1^{\mathrm{VB}},
%\label{S4-6}
%\end{equation}
%where $\Omega_1^{\mathrm{VB}}=d(M^{\mathrm{VB}}_1)^{T_\mathrm{V}}$. The constraint \eqref{S4-1-5} implies that $\Omega^{\mathrm{VB}}$ is a bipartite operator whose Schmidt number is at most $d'-1$, since the partial transpose won't change the Schmidt number. 
\emph{Alice's measurement is a full BSM.}-In this case, it is easy to see that since all of the (mutually orthogonal) projection operators can be transformed into each other via local unitary operations, i.e. $M^{\mathrm{VA}}_a=(U^{\mathrm{V}}_a \otimes \mathbb{I}^{\mathrm{A}})M^{\mathrm{VA}}_1 (U^{\mathrm{V}}_a \otimes \mathbb{I}^{\mathrm{A}})^\dagger$. This implies that for all $a$, 
\begin{equation}
\frac{1}{d}(\rho^{\mathrm{VB}}+r_a \tilde{\rho}_a^{\mathrm{VB}})=(U^{\mathrm{V}}_a \otimes \mathbb{I}^{\mathrm{A}})^{\dagger}(M^{\mathrm{VB}}_1)^{T_\mathrm{V}}(U^{\mathrm{V}}_a \otimes \mathbb{I}^{\mathrm{A}}), 
%(U^{\mathrm{V}}_a \otimes \mathbb{I}^{\mathrm{A}})\frac{1}{d}(\rho^{\mathrm{VB}}+r_1 \tilde{\rho}_1^{\mathrm{VB}})^{T_\mathrm{V}}(U^{\mathrm{V}}_a \otimes \mathbb{I}^{\mathrm{A}})^\dagger=M^{\mathrm{VB}}_a, 
\label{S4-7}
\end{equation}
Consequently, we have
\begin{equation}
    \sum^{d^2}_{a=1}\frac{1}{d}(\rho^{\mathrm{VB}}+r_a \tilde{\rho}_{N,a}^{\mathrm{VB}})=\sum^{d^2}_{a=1}(U^{\mathrm{V}}_a \otimes \mathbb{I}^{\mathrm{A}})^{\dagger}(M^{\mathrm{VB}}_1)^{T_\mathrm{V}}(U^{\mathrm{V}}_a \otimes \mathbb{I}^{\mathrm{A}})
    \label{S4-8}
\end{equation}
The left-hand-side of Eq. \eqref{S4-8} can be written as
\begin{equation}
    d\rho^{\mathrm{VB}}+\sum^{d^2}_{a=1}r_a \frac{\tilde{\rho}_{N,a}^{\mathrm{VB}}}{d}=d\rho^{\mathrm{VB}}+\sum^{d^2}_{a=1}r_a \tilde{N}^{VB}_a=d\rho^{\mathrm{VB}}+\mathbb{I}^{\mathrm{V}}\otimes r\tilde{\rho}^{\mathrm{B}}_N,
    \label{S4-9}
\end{equation}
where Eq. \eqref{S4-1-4} was used. Then by normalizing the coefficient of $\rho^{\mathrm{VB}}$, we arrive at
\begin{equation}
\rho^{\mathrm{VB}}+r\frac{\mathbb{I}^{\mathrm{V}}}{d}\otimes \tilde{\rho}^{\mathrm{B}}_N=\Omega^{\mathrm{VB}},
\label{S4-10}
\end{equation}
where $\Omega^{\mathrm{VB}}=(1/d)\sum^{d^2}_{a=1}(U^{\mathrm{V}}_a \otimes \mathbb{I}^{\mathrm{A}})^{\dagger}(M^{\mathrm{VB}}_1)^{T_\mathrm{V}}(U^{\mathrm{V}}_a \otimes \mathbb{I}^{\mathrm{A}})$. Since all $M^{\mathrm{VB}}_a$ has Schmidt number at most $d'-1$, i.e. satisfying Eq. \eqref{S4-1-6}, then $\Omega^{\mathrm{VB}}$ has Schmidt number at most $d'-1$, satisfying constraint \eqref{S4-2-4}. Consequently, with  $\rho^{\mathrm{VB}}=\rho^{\mathrm{AB}}$ and $\frac{\mathbb{I}^{\mathrm{V}}}{d}\otimes \tilde{\rho}^{\mathrm{B}}_N$ being a normalized separable operator, Eq. \eqref{S4-10} is equivalent with Eq. \eqref{S4-2-3}. We see that evaluating the generalized genuine $d'$-dimensional quantum teleportation robustness of data $\{\sigma_{a \mid \omega_x}^{\mathrm{B}}\}_{a, x}$ is equivalent to evaluating the generalized $d'$-dimensional entanglement robustness, which implies $R^{d^{\prime}}_{\text{tel}}=R^{d^{\prime}}_{\text{ent}}$.

\emph{Alice's measurement is a partial BSM.}-Suppose Alice's measurement has only two outcome $a=1,2$, and for $a=2$, the projection is $M^{\mathrm{VA}}_2=\mathbb{I}^{\mathrm{VA}}-M^{\mathrm{VA}}_1=\sum^{d^2}_{i=2}M^{\mathrm{VA}}_i$. In this case, 
\begin{equation}
\begin{aligned}
\frac{1}{d}(\rho^{\mathrm{VB}}+r_2 \tilde{\rho}_{N,2}^{\mathrm{VB}})=\sum^{d^2}_{i=2}(U^{\mathrm{V}}_i \otimes \mathbb{I}^{\mathrm{A}})^{\dagger}(M^{\mathrm{VB}}_1)^{T_\mathrm{V}}(U^{\mathrm{V}}_i \otimes \mathbb{I}^{\mathrm{A}})
\label{S4-11}
\end{aligned}
\end{equation}
From which it follows that 
\begin{equation}
    \sum^{2}_{a=1}\frac{1}{d}(\rho^{\mathrm{VB}}+r_a \tilde{\rho}_{N,a}^{\mathrm{VB}})=\sum^{d^2}_{i=1}(U^{\mathrm{V}}_i \otimes \mathbb{I}^{\mathrm{A}})^{\dagger}(M^{\mathrm{VB}}_1)^{T_\mathrm{V}}(U^{\mathrm{V}}_i \otimes \mathbb{I}^{\mathrm{A}}).
    \label{S4-12}
\end{equation}
By normalizing the coefficient of $\rho^{\mathrm{VB}}$, we arrive at
\begin{equation}
    \rho^{\mathrm{VB}}+\frac{d^2 r }{2}\frac{\mathbb{I}^{\mathrm{V}}}{d}\otimes \tilde{\rho}^{\mathrm{B}}_N=\Sigma^{\mathrm{VB}}.
    \label{S4-13}
\end{equation}
Where $\Sigma^{\mathrm{VB}}=(d/2)\sum^{d^2}_{a=1}(U^{\mathrm{V}}_a \otimes \mathbb{I}^{\mathrm{A}})^{\dagger}(M^{\mathrm{VB}}_1)^{T_\mathrm{V}}(U^{\mathrm{V}}_a \otimes \mathbb{I}^{\mathrm{A}})$. Since all $M^{\mathrm{VB}}_a$. Comparing Eq. \eqref{S4-13} with Eq. \eqref{S4-2-3}, we see that $R^{d^{\prime}}_{\text{tel}}=(2/d^2)R^{d^{\prime}}_{\text{ent}}$. Similarly, when the number of Alice's outcome is $o_A$, it is apparent to see that the relation is $R^{d^{\prime}}_{\text{tel}}=(o_A/d^2)R^{d^{\prime}}_{\text{ent}}$. In practical realization of program \eqref{S4-1} and \eqref{S4-2}, since $S_{d'-1}$ has a complicated structure, we choose to approximatively characterize $S_{d'-1}$ by its outer relaxation $S^{R}_{d'-1}$, i.e. $S_{d'-1}^R=\{M:(I_d \otimes \Lambda_{\frac{1}{d'-1}})(M) \geq 0\}$.. Similarly, the set $S_1$ can also be relaxed to the set of operators with positive partial transposition (PPT). Those relaxations have different impacts on $R^{d^{\prime}}_{\text{tel}}$ and $R^{d^{\prime}}_{\text{ent}}$, hence the practical ratio between the two quantities can vary case to case, especially when Alice perform partial BSM. Since $R^{d^{\prime}}_{\text{ent}}(\rho^{\mathrm{AB}})$ is non-null if and only if $\rho^{\mathrm{AB}}$ has Schmidt number at least $d'$, all $d'$-dimensional entangled states that have positive $\text{GHER}_{d'}$ can demonstrate genuine $d'$-dimensional QT. Notably, if the characterization of $S_{d'-1}$ is accurate, rather than its outer relaxation, all $d'$-dimensional entangled states can demonstrate genuine $d'$-dimensional QT.

Note that the condition (ii), i.e., the requirement of tomographically completeness is necessary for fairly evaluate the QT performance of given channel state. Relaxing this requirement would affect the value of $R^{d^{\prime}}_{\text{tel}}$. Since the underlying convex optimization program always searches for variables that maximize the objective function, relaxing the tomographically completeness, i.e., reducing the number of teleportation data, would results in decreasing the number of the constraints, which can only enlarge the feasible set. Hence, the obtained value of GHTR$_{d'}$ would always be less than or equal to the value computed under tomographic completeness. And this may change the certification results, e.g., the data that is actually produced by $d'$-dimensional entanglement1 can be falsely identified as accessible by ($d'-1$)-dimensional entanglement. Consequently, the assumption of tomographically complete inputs is a necessary condition, both in our study and other investigations of QT.

We summarize that, since the generalized genuine $d'$-dimensional quantum teleportation robustness $R^{d^{\prime}}_{\text{tel}}$ is proportional to the generalized $d'$-dimensional entanglement robustness $R^{d^{\prime}}_{\text{ent}}$, all channel states $\rho^{\mathrm{AB}}$ with $d'$-dimensional entanglement can demonstrate genuine $d'$-dimensional quantum teleportation.

Combined with the discussion in Appendix \ref{Ap:A}, while the previous methods for certifying genuine HDQT \cite{luo2019quantum,hu2020experimental} can be applied to determine whether the devices can teleport genuine HD quantum states, our criteria provide a much stricter certification that can only be satisfied by channel states whose entanglement dimension reach a certain level, thereby ensuring that the transmission capacity and noise resilience meet the required thresholds.

\begin{figure}[h!]
\centering
\includegraphics[width=10cm]{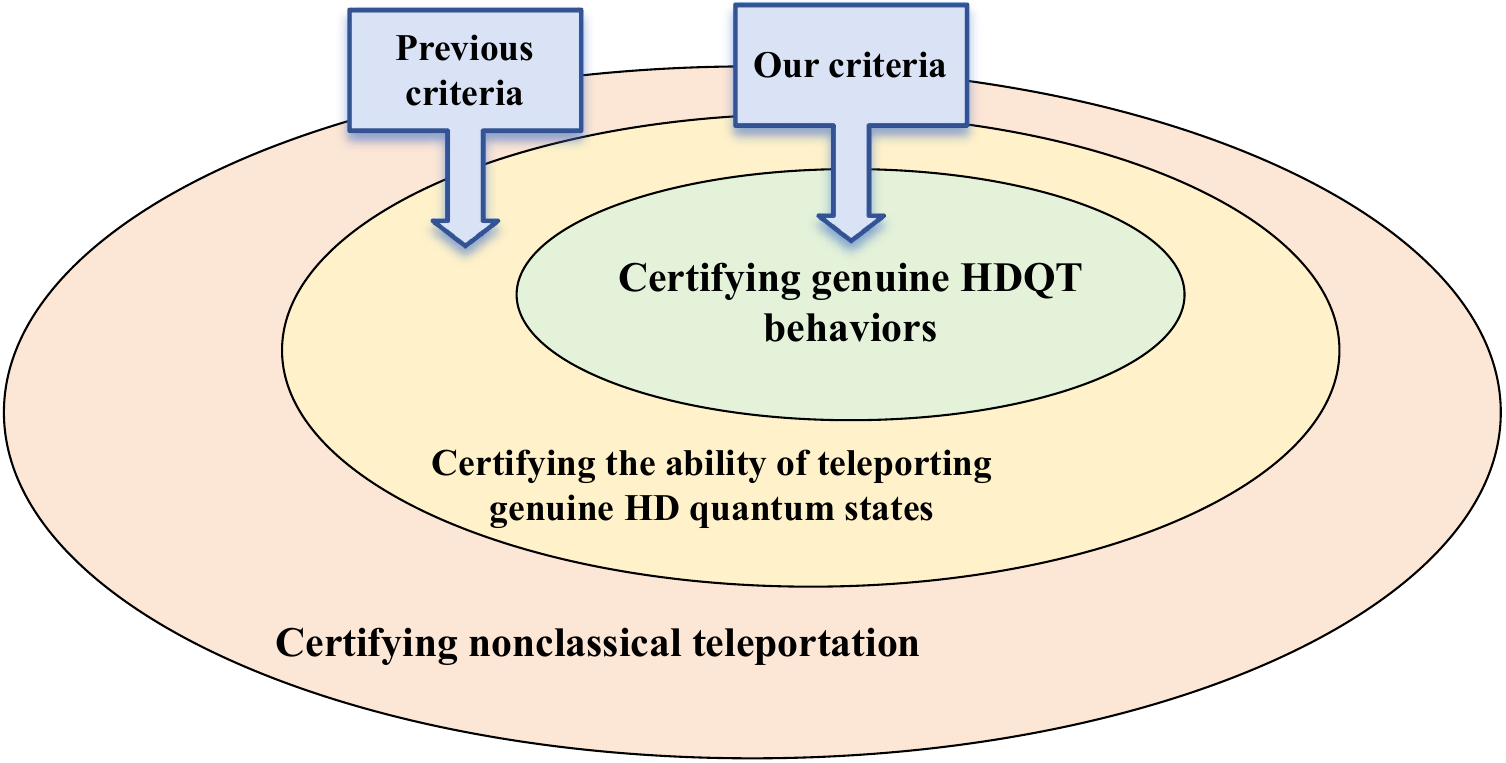}
\caption{The comparison of the stringency between our criteria and previous criteria. }
\label{comparison of certification ability}
\end{figure}

\end{document}